\documentclass[12pt]{article}
\usepackage{amsbsy,amsmath}
\usepackage{amsfonts}
\usepackage{amssymb}
\usepackage{amscd}
\usepackage{bbm}
\usepackage{fancybox}
\usepackage{cite}
\usepackage{amsmath,amsfonts,amsbsy}
\usepackage{pstricks,pst-node}
\usepackage[small,bf,hang]{caption2}
\usepackage[linktocpage]{hyperref}
\usepackage{graphicx}
\usepackage{epsfig}
\usepackage{psfrag}
\usepackage{comment}
\usepackage{xcolor}
\usepackage[normalem]{ulem}

\psset{unit=1.3cm,linewidth=.5pt,radius=.2}  

\usepackage{multirow}                     
\usepackage{float}                          
\usepackage{lscape}                         
\usepackage{bm}

\newcommand{\bb}{\bar\beta}

\newcommand{\beq}{\begin{equation}}
\newcommand{\eeq}{\end{equation}}
\newcommand{\bi}{\begin{itemize}}
\newcommand{\ei}{\end{itemize}}
\newcommand{\bt}{\begin{tabular}}
\newcommand{\et}{\end{tabular}}
\newcommand{\bc}{\begin{center}}
\newcommand{\ec}{\end{center}}

\newcommand{\be}{\begin{equation}}
\newcommand{\ee}{\end{equation}}
\newcommand{\bea}{\begin{eqnarray}}
\newcommand{\eea}{\end{eqnarray}}
\newcommand{\ba}{\begin{array}}
\newcommand{\ea}{\end{array}}

\def\bbox{{\,\lower0.9pt\vbox{\hrule \hbox{\vrule height 0.2 cm
\hskip 0.2 cm \vrule height 0.2 cm}\hrule}\,}}
\newcommand{\dsl}{\pa \kern-0.5em /}

\font\mybb=msbm10 at 12pt
\def\bb#1{\hbox{\mybb#1}}

\def\bR {\bb{R}}
\def\bE {\bb{E}}
\def\bB {\bb{B}}

\def\bH {\bb{H}}

\def\bA {\bb{A}}

\def\bI {\bb{I}}

\def\bfn{\mbox{\boldmath $\nabla$}}

\def\tr{{\rm tr}}

\makeatletter \@addtoreset{equation}{section} \makeatother

\def\slashchar#1{\setbox0=\hbox{$#1$}           
   \dimen0=\wd0                                 
   \setbox1=\hbox{/} \dimen1=\wd1               
   \ifdim\dimen0>\dimen1                        
      \rlap{\hbox to \dimen0{\hfil/\hfil}}      
      #1                                        
   \else                                        
      \rlap{\hbox to \dimen1{\hfil$#1$\hfil}}   
      /                                         
   \fi}

\begin{document}

\begin{titlepage}
\begin{center}

\vskip 1.5cm

{\Large \bf On p-form gauge theories and their conformal limits}

\vskip 1cm

{\bf Igor Bandos${}^1$, Kurt Lechner${}^2$, Dmitri Sorokin${}^2$ and Paul K.~Townsend${}^3$} \\

\vskip 25pt

{\em $^1$  \hskip -.1truecm
\em Department of Theoretical Physics, University of the Basque Country UPV/EHU, \\
P.O. Box 644, 48080 Bilbao, Spain, \\
and IKERBASQUE, Basque Foundation for Science,
48011, Bilbao, Spain. \vskip 5pt }

{email: {\tt igor.bandos@ehu.eus}} \\

\vskip .4truecm

{\em $^2$  \hskip -.1truecm
\em I.N.F.N., Sezione di Padova \\
and  Dipartimento di Fisica e Astronomia “Galileo Galilei”, \\
Universit\`a degli Studi di Padova, \\
Via F. Marzolo 8, 35131 Padova, Italy  \vskip 5pt }

{email: {\tt kurt.lechner@pd.infn.it, dmitri.sorokin@pd.infn.it}}
\vskip .4truecm

{\em $^3$ \hskip -.1truecm
\em  Department of Applied Mathematics and Theoretical Physics,\\ Centre for Mathematical Sciences, University of Cambridge,\\
Wilberforce Road, Cambridge, CB3 0WA, U.K.\vskip 5pt }

{email: {\tt P.K.Townsend@damtp.cam.ac.uk}} \\

\vskip .4truecm

\end{center}

\vskip 0.5cm
\begin{center} {\bf ABSTRACT}\\[3ex]
\end{center}

Relations between the various formulations of nonlinear p-form electrodynamics with
conformal-invariant weak-field and strong-field limits are clarified, with a focus on duality invariant $(2n-1)$-form electrodynamics and chiral $2n$-form electrodynamics in Minkowski spacetime of dimension $D=4n$ and $D=4n+2$,
respectively. We exhibit a new family of chiral 2-form electrodynamics in $D=6$ for which these
limits exhaust the possibilities for conformal invariance; the weak-field limit is related
by dimensional reduction to the recently discovered ModMax generalisation of Maxwell's equations. For $n>1$ we show
that the chiral `strong-field’  $2n$-form electrodynamics is related by dimensional reduction to
a new $Sl(2;\bR)$-duality invariant theory of $(2n-1)$-form electrodynamics.
\bigskip

\vfill

\end{titlepage}
\tableofcontents
\section{Introduction}

The Born-Infeld (BI) theory of non-linear electrodynamics in four dimensions (4D) and a related non-linear chiral 2-form electrodynamics in six dimensions (6D) arise
naturally in String/M-theory as truncations of the effective low-energy dynamics of the D3-brane and M5-brane, respectively. They are related by a consistent dimensional-reduction/truncation inherited from the fact that the  D3-brane effective (4D worldvolume) action is a consistent dimensional reduction of the M5-brane effective (6D worldvolume) action \cite{Berman:1998va,Nurmagambetov:1998gp}, where `consistent' means that any solution of the lower-dimensional equations `lifts' to a solution of the higher-dimensional equations.  For this reason, it is useful to consider the 4D BI theory and the 6D chiral 2-form electrodynamics on the M5-brane as partners in what we shall call the ``D3/M5 pair''.

In this paper we explore the possibilities for other 4D/6D pairs in the context of various formulations of both the 4D and 6D partners. In the 4D case we consider the generic nonlinear electrodynamics
theory that is both Lorentz invariant and invariant under  an $SO(2)$ electromagnetic-duality group (as is the BI theory).  In the 6D case we consider the general Lorentz invariant nonlinear chiral 2-form electrodynamics; as for the D3/M5 pair, 6D chirality implies electromagnetic-duality of the 4D nonlinear electrodynamics obtained by a consistent reduction/truncation.

Our principal new result is  a one-parameter generalization of the D3/M5 pair for which the 4D partner is the ``generalized BI electrodynamics'' of  \cite{Bandos:2020jsw}, which has the property that its weak-field and strong-field limits exhaust the possibilities for conformal duality-invariant 4D electrodynamics. The 6D partner is a new interacting familiy of chiral 2-form electrodynamics theories with the same property: its weak-field and strong-field limits exhaust the possibilities for conformal chiral 2-form electrodynamics. For the 4D case it was shown in \cite{Bandos:2020jsw} that the strong-field limit is the same as that of the BI theory, i.e. the $Sl(2;\bR)$-duality invariant Bialynicki-Birula (BB) electrodynamics \cite{BialynickiBirula:1984tx}, but the weak-field limit is a novel interacting one-parameter ``ModMax'' generalization of  Maxwell electrodynamics. Here we find that its 6D partner has the same strong-field limit as that of the `M5' chiral 2-form electrodynamics \cite{Gibbons:2000ck,Townsend:2019ils} but its weak-field limit is a new conformal chiral 2-form electrodynamics that contains
ModMax electrodynamics as a consistent reduction/truncation.

We also explore the possibility of higher-dimensional analogs of these 4D/6D pairs. A class of duality invariant BI-inspired $(2n-1)$-form electrodynamics theories in a (Minkowski) spacetime of dimension
$D=4n$ was found in \cite{Gibbons:1995cv}.  A natural question  is whether they are obtainable by consistent truncation/reduction of some chiral $2n$-form electrodynamics in $D=4n+2$ for $n>1$. For $n=2$ (at least) the answer is no, because the leading (quartic) interaction terms in the weak-field expansion  do not match those found by reduction/truncation of the quartic interaction terms in  the weak-field expansion of any 10D chiral 4-form electrodynamics that has a weak-field expansion \cite{Buratti:2019cbm,Buratti:2019guq}.
This result leaves open the possibility that the higher-dimensional generalization of BB electrodynamics found in \cite{Chruscinski:2000zm} from the strong-field limit of the BI-type theory of  Gibbons and Rasheed
\cite{Gibbons:1995cv} is a consistent truncation of some analogous higher-dimensional generalization of  the strong-field `M5' chiral 2-form electrodynamics.

In fact, there is a natural generalization of the strong-field `M5' electrodynamics to an  interacting conformal chiral $2n$-form electrodynamics in  $D=4n+2$ for any $n>1$ \cite{Townsend:2019koy} and we find here, by reduction/truncation, the corresponding conformal duality-invariant $(2n-1)$-form electrodynamics in $D=4n$; for $n=1$ this is BB electrodynamics but for $n>1$ it is a new higher-dimensional generalization that differs from that of \cite{Chruscinski:2000zm}. This is possible because for $n>1$ the requirements of conformal invariance and $Sl(2;\bR)$-duality invariance do not determine the Hamiltonian density uniquely.

Another purpose of this paper is to explore the relations between the various formulations of the generic p-form electrodynamics theories mentioned above, mostly focusing on the 4D and 6D cases. In the 4D case, we have the standard Lagrangian and Hamiltonian formulations of generic nonlinear duality-invariant (1-form) electrodynamics but in neither formulation are both Lorentz invariance and duality invariance manifest. In the 6D case we have a Hamiltonian formulation \cite{Henneaux:1988gg,Bergshoeff:1998vx,Gibbons:2000ck,Chruscinski:2000zm,Townsend:2019ils}, which we develop further here. However, because of chirality \cite{Marcus:1982yu}, the closest one can get to a standard Lagrangian formulation is what we call here the Perry-Schwarz formulation in which only a 5D Lorentz invariance is manifest \cite{Perry:1996mk}; this is essentially a variant of the Hamiltonian formulation for which the manifest Lorentz symmetry subgroup is $SO(1,4)$ rather than $SO(5)$.  For both 4D and 6D there {\it is} a formulation  in which all symmetries are manifest; this is the PST formulation, which involves an additional closed 1-form field \cite{Pasti:1995tn,Pasti:1996vs,Maznytsia:1998xw}, but an additional non-manifest gauge
invariance is then needed for equivalence with the `standard' formulations.

A consequence of the fact that not all symmetries (and gauge-invariances) can be simultaneously manifest is that there is always some symmetry (or gauge invariance) that must be imposed.  This means that the function defining the particular model (e.g. Hamiltonian density in the Hamiltonian formulation) must satisfy a condition, and in each case
this can be expressed as a partial differential equation (PDE) in two independent variables (e.g. two independent rotation scalars for the Hamiltonian density). In the 4D case this PDE has been found various times using various methods \cite{Gaillard:1981rj,BialynickiBirula:1984tx,Gibbons:1995cv,Deser:1997gq}, and the same is true for 6D \cite{Perry:1996mk,Bekaert:1998yp,Bekaert:1999ue}. For an appropriate choice of the two independent variables in each case, the PDE is the same in all cases: not only do different formulations of the 4D or 6D theories lead to the same PDE, but also (as observed in \cite{Perry:1996mk}) one finds the same PDE for both 4D and 6D. Here we explain this result by showing how the 6D PDE is mapped into the 4D PDE by the process of consistent reduction/truncation. This implies a one-to-one correspondence between 4D nonlinear theories of duality-invariant electrodynamics and 6D nonlinear chiral 2-form electrodynamics. 

We shall begin with a review of the $p=0,1,2$ cases of $p$-form electrodynamics in a Minkowski spacetime of dimension $D=2p+2$; these are the dimensions that allow either electromagnetic duality invariance (for odd $p$) or chirality (for even $p$). This review includes some new material; for example we recover by Hamiltonian methods the result of \cite{Buratti:2019guq} that any chiral 0-form electrodynamics is a free-field theory, and we give another derivation of the condition on the Hamiltonian density for Lorentz invariance of the generic 6D chiral 2-form electrodynamics. We also explain how unusual features of the
Legendre transform for BB-electrodynamics do not prevent an equivalence of the Hamiltonian and Lagrangian formulations even though the (canonical) Lagrangian density is identically zero.  These ``preliminaries'' are followed by an exposition of the PST method in which some details passed over in earlier works on this topic are explained.

The abstract `universal' PDE that one must solve to find any particular pair of duality invariant 4D electrodynamics and chiral 2-form 6D electrodynamics has a known general solution \cite{C-H} but its application to electrodynamics (see e.g. \cite{Gibbons:1995cv,Perry:1996mk,Hatsuda:1999ys}) requires additional physical constraints, such as the requirement of an `acceptable' weak-field limit. It has generally been supposed that the weak-field
limit must be a free-field theory, but it was argued in \cite{Bandos:2020jsw} that the ModMax theory mentioned above is a physically acceptable 
alternative weak-field limit for 4D electrodynamics; here we give some details of the derivation of the one-parameter generalization of
BI theory that has ModMax as its weak-field limit. The principal novelty is its 6D analog (and its weak-field limit) which is the new family of chiral 2-form electrodynamics advertised in our Abstract; we also provide an alternative proof of conformal invariance of the 
weak-field and strong-field limits that applies both in 4D and 6D. 

We then turn to the higher-dimension $p$-form theories, using their PST formulation
to obtain the results mentioned above. In addition, we discuss the Legendre transform
for a class of (odd-$p$) duality-invariant $p$-form electrodynamics theories that include
the ``generalized BI'' theory of \cite{Bandos:2020jsw}. There we showed how the weak-field limits are related by a Legendre transform; here we use a more powerful method that not only avoids the need to take the weak-field limit but also applies for any odd $p\ge1$.

We conclude with an overview of the paper and
some discussion of open problems.

\section{p-form electrodynamics: preliminaries}

By ``p-form electrodynamics'' we mean here an abelian gauge-field theory for a $p$-form potential $A$ with $(p+1)$-form
field-strength $F=dA$,  in a Minkowski spacetime of dimension $2p+2$. For $p$ odd, $F$ transforms irreducibly with
respect to the Lorentz group. In contrast,   $F=F_+ \oplus F_-$ for even $p$, where $F_{\pm}$ are the (anti)self-dual components of $F$, which transform
as distinct irreducible representations\footnote{
We may also write $F= F_++F_-$ for odd p but then $F_-=F_+^*$ for real $F$,  so we do not have
a decomposition of $F$ into irreducible representations of the Lorenz group.} that are exchanged by parity; in this case we may set $F=F_+$ to get a ``chiral'' theory.
For any $p$ there could be $(p-1)$-brane sources, which may be `electric' or magnetic' (or `dyonic') for odd $p$, but we consider here
only source-free theories.

We also assume that the Lagrangian density is an ultralocal  Lorentz-scalar  function of $F$; i.e. it does not involve any
derivatives of $F$. This means that the dimension of the space of initial conditions at any space point is unaffected by
interactions; in other words, interactions  allowed by this assumption do not change the number of degrees of freedom.
In the context of the Hamiltonian formulation, this implies that the canonical structure is unaffected by the interactions.

The lowest odd $p$ is $p=1$; the free field case is Maxwell electrodynamics and the best-known
interacting example is Born-Infeld electrodynamics. The lowest even $p$ is $p=0$; the free field
case is the ``chiral boson''. Self-interactions are not possible for $p=0$ \cite{Buratti:2019guq}; here we give another proof of this statement.  The next-to-lowest even $p$ is $p=2$, and this includes not only the free chiral 2-form electrodynamics but also the BI-type theory on the M5-brane.

As mentioned in the introduction, the $p=1$ and $p=2$ cases are linked by dimensional
reduction, so it is convenient to consider them together. The main aim of this section is to present the basic properties of these theories from both a Lagrangian and a Hamiltonian
perspective; much of this will be review but some known results are recovered by different methods. Our aim is to exhibit the unity underlying the different formulations of p-form electrodynamics subject to a duality-invariance/chirality restriction. 

\subsection{2D chiral 0-form electrodynamics}

In this case $F=d\varphi$ for a scalar field $\varphi(\tau,\sigma)$. The chirality condition on $F$ is
$\dot\varphi = \varphi'$ (with $\dot\varphi\equiv\partial_\tau\varphi$ and $\varphi'\equiv\partial_\sigma\varphi$), which is the standard chiral boson equation. However, to address the issue of
possible interactions in a way that is in line with the general definition above of p-form electrodynamics
it is convenient to start from a Hamiltonian formulation in which the Hamiltonian field equations follow
from variation of the  phase-space (or canonical) action
\begin{equation}\label{CBact}
I= \int \! dt \int\! d\sigma \left\{\mp \dot\varphi\, \psi - {\cal H}(\psi)\right\} \, , \qquad \psi=\varphi'\, ,
\end{equation}
where an overdot indicates a time derivative and a prime indicates a space derivative. We allow the Hamiltonian density to be
an arbitrary function of  the `magnetic' field $\psi$, which is invariant under the  `semi-local'  gauge transformation $\varphi \to \varphi + \alpha(t)$, where $\alpha$ is an arbitrary function of time.  The field equation is
\begin{equation}\label{fe}
\dot\varphi' = \mp  \frac12 ({\cal H}_\psi)'\, .
\end{equation}
The action (\ref{CBact}) is manifestly invariant under translations in time and space, and the corresponding Noether charges are
\begin{equation}
H= \int\! d\sigma \, {\cal H}\, , \qquad P = \pm\int\! d\sigma \, \psi^2 \, .
\end{equation}
These are time-independent (for suitable boundary conditions) since the field equation implies that
\begin{equation}\label{translation}
\partial_t {\cal H} = \mp \frac14({\cal H}_\psi^2)' \, , \qquad \partial_t[\psi^2]=  \mp({\cal H} - \psi{\cal H}_\psi )' \, .
\end{equation}

If the action (\ref{CBact}) is Lorentz invariant  then  the Lorentz boost Noether charge is
\begin{equation}
L = tP - \int \! d\sigma \, \sigma {\cal H}\, .
\end{equation}
However, we must check that this is a conserved charge because the action is not manifestly Lorentz invariant.
Using the field equation (\ref{fe}), and assuming that boundary terms are zero, we have
\begin{equation}
\dot L = P - \int \! d\sigma\, \sigma \dot {\cal H} = \pm \int\! d\sigma \left(\psi- \frac12 {\cal H}_\psi\right)\left( \psi + \frac12{\cal H}_\psi\right)\, ,
\end{equation}
which is zero iff ${\cal H}_\psi =\pm 2\psi$. We may assume without loss of generality that ${\cal H}$ is non-negative, and zero for $\psi=0$, in which case
\begin{equation}
{\cal H}(\psi) = \psi^2 \, .
\end{equation}
We thus conclude that the most general Lorentz invariant action of the form (\ref{CBact}) is
\begin{equation}\label{FJ}
I= \int \! dt \int\! d\sigma\,  (\mp \dot\varphi - \varphi')\varphi'\, .
\end{equation}
This is the Floreanini-Jackiw action for a free (anti)chiral boson  \cite{Floreanini:1987as}.
In agreement with \cite{Buratti:2019guq} we conclude that no Lorentz invariant self-interactions are possible.

\subsection{4D nonlinear 1-form electrodynamics}\label{subsec:4D}

We now review aspects of nonlinear theories of electrodynamics in a 4D Minkowski spacetime. A convenient starting point is
the phase-space action
\begin{equation}\label{4Dact}
I= \int\! dt\int\! d^3 \sigma \left\{ {\bf E}\cdot{\bf D} - {\cal H}({\bf D}, {\bf B})  \right\}\, ,
\end{equation}
where $(t,\boldsymbol{\sigma})$ are the time and cartesian space coordinates. The electric field ${\bf E}$ and magnetic induction field ${\bf B}$ are
defined as
\begin{equation}
{\bf E} =  \boldsymbol{\nabla} A_0 -\dot{\bf A} \, , \qquad {\bf B} = \boldsymbol{\nabla}\times {\bf A} \, ,
\end{equation}
which means (in this context)  that the electric-displacement $3$-vector ${\bf D}$ is canonically conjugate to minus the vector potential ${\bf A}$,
while the scalar potential $A_0$ is a Lagrange multiplier for the constraint $\boldsymbol{\nabla}\cdot {\bf D}=0$.  If we choose the Hamiltonian density
to be a function of the three independent rotation scalars
\begin{equation}\label{BBvars}
s = \frac12(  |{\bf D}|^2 + |{\bf B}|^2) \, , \qquad \xi = \frac12(  |{\bf D}|^2 - |{\bf B}|^2)\, , \qquad \eta = {\bf D}\cdot{\bf B} \, ,
\end{equation}
then the action is invariant under time and space translations, and rotations. It is also Lorentz invariant if \cite{BialynickiBirula:1984tx}
\begin{equation}\label{BB-Lorentz}
{\cal H}^2_s -{\cal H}^2_\xi - {\cal H}^2_\eta =1\, .
\end{equation}

The Hamiltonian density is further invariant under an $SO(2)$ `duality' rotation of the 2-vector with $3$-vector components $({\bf D},{\bf B})$ if it is a function only of $s$ and
\begin{equation}
p= |{\bf D}  \times {\bf B}| \, ,
\end{equation}
which is related to $(s,\xi,\eta)$ by
\begin{equation}\label{speq}
s^2-p^2 = \xi^2 + \eta^2\, .
\end{equation}
In terms of these variables the condition for Lorentz invariance \eqref{BB-Lorentz} becomes
\begin{equation}\label{PDEsp}
{\cal H}^2_s + \frac{2s}p{\cal H}_s{\cal H}_p + {\cal H}^2_p  =1 \, .
\end{equation}
As $p$ has an $Sl(2;\bR)$ electromagnetic duality invariance, the same will be true
of ${\cal H}$ iff it is a function only of $p$, and then Lorentz invariance requires
${\cal H} = \pm p$.  Choosing the positive sign, we have the interacting conformal
electrodynamics of  Bialynicki-Birula \cite{BialynickiBirula:1984tx}
\be\label{BBd4}
{\cal H}_{BB} =|\mathbf D\times \mathbf B|\,.
\ee

An alternative basis for $SO(2)$ duality invariant rotation scalars is
\begin{equation}\label{uv4D}
{\tt u} = \frac12(s + \sqrt{s^2-p^2})\, , \qquad {\tt v}= \frac12(s - \sqrt{s^2-p^2}) \, .
\end{equation}
These new variables are well defined since it follows from (\ref{speq}) that
$s^2-p^2\ge0$. In terms of them, the Lorentz-invariance condition (\ref{PDEsp}) simplifies to
\begin{equation}\label{BB-Lorentz2}
{\cal H}_{\tt u} {\cal H}_{\tt v} =1\, .
\end{equation}

Remarkably, an equation formally equivalent to (\ref{BB-Lorentz2}) appears in the manifestly Lorentz-invariant Lagrangian formulation as the condition for electromagnetic duality
invariance \cite{Gaillard:1981rj,Gibbons:1995cv,Gaillard:1997zr,Gaillard:1997rt,Hatsuda:1999ys,Kuzenko:2000tg,Bekaert:2001wa,Ivanov:2002ab,Ivanov:2003uj,Kuzenko:2019nlm}. The Lagrangian density ${\cal L}$ of a general Lorentz invariant theory of electrodynamics may be written in terms of the two independent Lorentz scalars
\begin{equation}\label{SPinv}
S = -\frac 14 F_{\mu\nu}F^{\mu\nu}=\frac12( |{\bf E}|^2 - |{\bf B}|^2) \, , \qquad P =
-\frac 18 \varepsilon^{\mu\nu\lambda\rho}F_{\mu\nu}F_{\lambda\rho}={\bf E} \cdot {\bf B} \,.
\end{equation}
In the Lagrangian formulation,  the electric-displacement vector ${\bf D}$ is defined as  $\partial {\cal L}/\partial {\bf E}$, and the condition for electromagnetic-duality invariance of the EL equations of a Lagrangian density ${\cal L}(S,P)$ is
\begin{equation}\label{PDESP}
 {\cal L}_S^2 - \frac{2S}{P}{\cal L}_S {\cal L}_P - {\cal L}_P^2 =1\, ,
\end{equation}
which is very similar to (\ref{PDEsp}). In terms of the new variables
\begin{equation}\label{dual-inv}
U = \frac12 \left( S -  \sqrt{S^2 +P^2}\right) \, ,  \qquad  V = \frac12 \left(S + \sqrt{S^2 + P^2}\right) \, ,
\end{equation}
this duality-invariance condition simplifies to \cite{Gibbons:1995cv}
\begin{equation}\label{GibRash}
{\cal L}_U {\cal L}_V =1\, ,
\end{equation}
which is formally identical to (\ref{BB-Lorentz2}).

To summarise, the Hamiltonian and Lagrangian densities of a generic non-linear Lorentz and duality invariant
electrodynamics theory are both solutions of the same PDE for particular choices of the two pairs of variables
on which they depend. For any ${\cal H}({\tt u, v})$ satisfying (\ref{BB-Lorentz2}) its Legendre transform
(with respect to ${\bf D}$) will be some ${\cal L}(U,V)$ satisfying (\ref{GibRash}) \cite{Deser:1997gq}.
For example, one solution to the equation (\ref{BB-Lorentz2}) is
\begin{eqnarray}\label{cHDBI=}
{\cal H}_{(T)} &=&  \sqrt{T^2 +2T({\tt u+v}) + 4{\tt uv}} -T \nonumber \\
&=& \sqrt{T^2 + T ( |{\bf D}|^2 + |{\bf B}|^2) + |{\bf D}\times {\bf B}|^2} -T \, ,
\end{eqnarray}
which is the BI Hamiltonian density.  The corresponding solution to (\ref{GibRash}) is
\begin{eqnarray}
{\cal L}_{(T)} &=& T- \sqrt{T^2 - 2T(U+V) + 4UV}  \nonumber \\
&=&  T- \sqrt{T^2 - T(|{\bf E}|^2 - |{\bf B}|^2)  - ({\bf E}\cdot {\bf B})^2 } \, ,
\end{eqnarray}
which is the BI Lagrangian density.

\subsubsection{BB electrodynamics and the Legendre transform}

Starting with the BI Hamiltonian density of (\ref{cHDBI=}) we may, following Bialynicki-Birula \cite{BialynickiBirula:1984tx},
take the $T\to0$ limit to arrive at the BB Hamiltonian density of (\ref{BBd4}):
\begin{equation}\label{bbham}
\lim_{T\to0} {\cal H}_{(T)}({\tt u,v}) = 2\sqrt{\tt uv} = |{\bf D} \times {\bf B}| = {\cal H}_{BB}\, .
\end{equation}
As $T$ has dimensions of energy density, this limit should be seen as a strong-field limit in which the energy density is much greater than $T$,
just as the weak-field limit should be seen as the limit in which the energy density is much less than $T$. One might expect these limits to yield
conformal invariant theories, as is evidently true for the weak field limit since the Hamiltonian equations in this limit are Maxwell's equations. As shown
in  \cite{BialynickiBirula:1984tx}, it is also true for the strong-field limit, which has the additional feature that  ${\cal H}_{BB}$ is invariant under an
$Sl(2;\bR)$ electromagnetic-duality acting on  the 3-vector-valued 2-vector $({\bf D},{\bf B})$. 

Let us find the  Lagrangian density of this BB-electrodynamics by taking the Legendre transform of ${\cal H}_{BB}$; the first step is to define
\begin{equation}\label{E-BB}
{\bf E} := \frac{\partial\cal H_{BB}}{\partial {\bf D}} = - {\bf n} \times {\bf B}\, , \qquad {\bf n} = \frac{{\bf D}\times {\bf B}}{|{\bf D}\times {\bf B}|}\, .
\end{equation}
The `canonical' BB Lagrangian density is then defined as
\begin{equation}\label{DefLT}
{\cal L}_{BB}({\bf E},{\bf B}) := \sup_{\bf D}\, [ {\bf D}\cdot {\bf E} - {\cal H}_{BB}]\,.
\end{equation}
In principle, this requires us to find the ${\bf D}$ that maximises the
expression in brackets. However, in this case
\begin{equation}
{\bf D}\cdot {\bf E} - {\cal H}_{BB} = {\bf n} \cdot ({\bf D} \times {\bf B}) - |{\bf D} \times {\bf B}| \equiv 0\, ,
\end{equation}
so we conclude, following \cite{BialynickiBirula:1984tx}, that
\begin{equation}
 {\cal L}_{BB}({\bf E},{\bf B}) \equiv 0\, .
\end{equation}

Despite this conclusion it remains true that the Legendre transform of ${\cal L}_{BB}$ is ${\cal H}_{BB}$. We shall provide an explicit proof below, but
we first wish to point out that it is a consequence of the convexity of ${\cal H}_{BB}$ as a function of ${\bf D}$ since a general theorem (see e.g. \cite{Arnold:1989}) guarantees that
the Legendre transform defined as in (\ref{DefLT}) (but now with ${\cal L}_{BB}$ and ${\cal H}_{BB}$ exchanged) is an involution when acting on convex functions. So
all we really need to prove is convexity of ${\cal H}_{BB}$ and we can do this by investigating its Hessian matrix:
\begin{equation}
\frac{\partial^2{\cal H}_{BB}} {\partial D^i \partial D^j} = \frac{|{\bf B}|^2}{{\cal H}_{BB}}\,  n_i n_j\, .
\end{equation}
As all eigenvalues of this matrix are non-negative for all ${\bf D}$, given any ${\bf B}$, the function ${\cal H}_{BB}$ is convex. This is sufficient for the theorem,  but the
case under consideration is special because ${\cal H}_{BB}$ is not strictly
convex (some eigenvalues of its Hessian are zero) and a consequence of this is that we cannot solve unambiguously for ${\bf D}$ the equation defining ${\bf E}$, i.e. (\ref{E-BB}). A corollary is that the equation defining ${\bf E}$ imposes constraints on ${\bf E}$; from (\ref{E-BB}) we see that these constraints are\footnote{The $T\to0$ limit of the BI Lagrangian
density may be taken if $P=0$ is imposed, with the result that
${\cal L}_{BB}=0$, but this attempt to take the strong-field limit misses the $S=0$ constraint.}
\begin{equation}
    S=P=0\, ,
\end{equation}
where $(S,P)$ are  the Lorentz scalars of (\ref{SPinv}). In other words, although ${\cal L}_{BB}$ is identically zero the domain of the ${\bf E}$ field-space in which it is defined is restricted by $S=P=0$. This means that the Legendre transform of ${\cal L}_{BB}$ is
\begin{equation}\label{LTL}
{\cal H}({\bf D}, {\bf B}) = \sup_{\{{\bf E}| S=P=0\}}
[ {\bf D}\cdot {\bf E} - 0]\, .
\end{equation}
Now we have a constrained variational problem: we must find the ${\bf E}$, within its allowed domain, that maximises the expression in brackets.

We can solve this constrained variational problem by the Lagrange multiplier method;  i.e. we first look for the stationary points of
\begin{equation}\label{constrainedH}
{\cal H}( {\bf D}, {\bf B}; {\bf E}, \lambda,\mu) := {\bf D}\cdot {\bf E} - \lambda S - \mu P\, ,
\end{equation}
where we must vary ${\bf E}$ (now without constraints) and the Lagrange multiplier fields $(\lambda,\mu)$. We must then examine the results of this calculation to
find the maximum of ${\bf D}\cdot {\bf E}$, rather than a minimum or some other
stationary value.

Varying the function defined in (\ref{constrainedH}) with respect to ${\bf E}$ we have ${\bf E}  = \lambda^{-1} ({\bf D}-\mu {\bf B})$, and  back-substitution yields
 \begin{equation}
{\cal H}( {\bf D}, {\bf B};  \lambda,\mu) = \frac{1}{2\lambda} |{\bf D} - \mu{\bf B}|^2 + \frac{\lambda}{2} |{\bf B}|^2\, .
 \end{equation}
Varying this with respect to $\mu$ we have $\mu |{\bf B}|^2= {\bf D}\cdot {\bf B}$ and back-substitution yields
\begin{equation}
{\cal H}( {\bf D}, {\bf B};  e)= \frac12 \left[ e^{-1} |{\bf D}\times {\bf B}|^2 +e \right] \, , \qquad e= \lambda|{\bf B}|^2 \, .
\end{equation}
Finally, elimination of $e$ yields $e= \pm  |{\bf D}\times {\bf B}|$ and hence
\begin{equation}
{\cal H}( {\bf D}, {\bf B})  = \pm |{\bf D}\times {\bf B}|\, ,
 \end{equation}
for unrestricted ${\bf D}$.  The maximum is achieved by choosing the top sign, so the Legendre transform of ${\cal L}_{BB}$ is ${\cal H}_{BB}$, as claimed.

The unusual aspect  of this particular pairing of convex functions by the Legendre transform is that the information in ${\cal L}_{BB}$ that is needed for the reconstruction of ${\cal H}_{BB}$ resides entirely in the restriction on its domain, not on its values within this domain!

Finally, we should point out that the effect of using the Lagrange multiplier method to solve the constrained variation problem,   implicit in the definition
(\ref{LTL}) of ${\cal H}_{BB}$ as the Legendre transform of ${\cal L}_{BB}$, is to replace the identically zero canonical Lagrangian density by its ``weak''
equivalent\footnote{In  the terminology of constrained Hamiltonian systems \cite{Dirac:1950pj}.}
\begin{equation}
{\cal L}_{BB} \approx \lambda S + \mu P\, .
\end{equation}
This is precisely the Lagrangian density proposed in \cite{BialynickiBirula:1992qj}, where it was verified that the field equations are equivalent to the
Hamiltonian field equations found from ${\cal H}_{BB}$.  Now we can see more precisely why this is true.

\subsubsection{Off-shell duality invariance}\label{off-shell}

For the phase-space action (\ref{4Dact}), any invariance of the Hamiltonian density will be an invariance of the
Hamiltonian field equations but not necessarily of the action itself; if the action is not invariant we have an ``on-shell'' symmetry. Electromagnetic duality invariance, acting as a $SO(2)$ transformation on $({\bf D},{\bf B})$, is an example; it is an ``on-shell'' symmetry
when ${\cal H}$ is duality invariant but it cannot be an ``off-shell'' symmetry of the action (\ref{4Dact}) because
${\bf B}=\boldsymbol{\nabla}\times {\bf A}$ is identically divergence-free but its duality partner ${\bf D}$ is divergence-free only as the result of a constraint imposed by a Lagrange multiplier.  However, this ``off-shell'' difference between ${\bf D}$ and ${\bf B}$ may be eliminated, in the absence of sources, by solving the constraint on ${\bf D}$ in terms of a new `dual' vector potential $\tilde {\bf A}$:
\begin{equation}
{\bf D} = \boldsymbol{\nabla}\times \tilde {\bf A}\,.
\end{equation}
This replacement converts the action (\ref{4Dact}) into one that is a functional of the pair of vector potentials
$({\bf A},\tilde{\bf A})$:
\begin{equation}\label{AtAaction}
I[{\bf A}, \tilde{\bf A}] = \int\! dt\int\! d^3 \sigma \left\{ -\dot{\bf A} \cdot \boldsymbol{\nabla} \times \tilde{{\bf A}}  -
{\cal H}({\bf D}, {\bf B})  \right\}\, .
\end{equation}
We now have both  $\boldsymbol{\nabla}\cdot{\bf D}=0$  and
$\boldsymbol{\nabla}\cdot{\bf B}=0$ as identities, while variation with respect to ${\bf A}$ and $\tilde{\bf A}$ yields the remainder of the  ``macroscopic Maxwell equations'':
\begin{equation}\label{macroMax}
\dot {\bf B} = -\boldsymbol{\nabla} \times {\bf E}\, ,\qquad  \dot{\bf D} =  \boldsymbol{\nabla} \times {\bf H}\, ,
\end{equation}
where $({\bf E},{\bf H})$ are again given by the constitutive relations:
\begin{equation}\label{constit}
    {\bf E} = \frac{\partial {\cal H}}{\partial {\bf D}} \, , \qquad {\bf H} = \frac{\partial {\cal H}}{\partial {\bf B}}\, .
\end{equation}
Lorentz invariance is not guaranteed; the condition for it is \cite{BialynickiBirula:1984tx}
\begin{equation}\label{Linvcon}
{\bf E} \times {\bf H} = {\bf D}\times {\bf B} \, .
\end{equation}
Electromagnetic duality invariance is not guaranteed either but it now acts as an $SO(2)$ rotation of the vector-potential doublet $({\bf A},\tilde{\bf A})$, and is a symmetry of the action (\ref{AtAaction}) if ${\cal H}$ is an $SO(2)$ invariant; equivalently, if \cite{BialynickiBirula:1984tx}
\begin{equation}\label{Dinvcon}
{\bf E}\cdot {\bf B} = {\bf H} \cdot {\bf D}\, .
\end{equation}
The relations (\ref{Linvcon}) and (\ref{Dinvcon}) are jointly equivalent to the PDE  (\ref{PDEsp}) to be satisfied by ${\cal H}$ and, as expected,
\eqref{Dinvcon} is an identity when ${\cal H}={\cal H}(s,p)$.
An advantage of this formulation is that duality is now an off-shell symmetry. Noether's theorem therefore applies and there is a corresponding conserved Noether charge \cite{BialynickiBirula:1984tx,Deser:1976iy}. This method was used in \cite{Mezincescu:2019vxk} to construct an $Sl(2;\bR)$-invariant action for Bialynicki-Birula electrodynamics and thereby find expressions for its $Sl(2;\bR)$ triplet of conserved Noether charges.

If we further change notation by setting $(\tilde {\bf A}, {\bf A})=({\bf A}^1,{\bf A}^2)$, then
\begin{equation}\label{EBcorresp}
 ({\bf D}, {\bf B})=({\bf B}^1,{\bf B}^2) \, , \qquad
 (-{\bf H},  {\bf E})=({\bf E}^1, {\bf E}^2)  \, ,
\end{equation}
where, for ${\tt a}=1,2$,
\begin{equation}\label{defsEB}
    {\bf E}^{\tt a} =  \boldsymbol{\nabla} A^{\tt a}_0 - \dot{\bf A}^{\tt a}\, ,
    \qquad {\bf B}^{\tt a} = \boldsymbol{\nabla} \times {\bf A}^{\tt a}\, .
\end{equation}
Notice that these fields are invariant under the following gauge transformations with an $SO(2)$-doublet of scalar parameters $\alpha^{\tt a}$:
\begin{equation}\label{gaugetrans}
    A_0^{\tt a} \to A_0^{\tt a} + \dot \alpha^{\tt a}\, , \qquad {\bf A}^{\tt a}\to {\bf A}^{\tt a} + \boldsymbol{\nabla} \alpha^{\tt a}\, .
\end{equation}
Notice too that the definitions (\ref{defsEB}) imply the identities
\begin{equation}\label{fromdefs}
    \dot {\bf B}^{\tt a} \equiv - \boldsymbol{\nabla} \times {\bf E}^{\tt a}\, ,
    \qquad \boldsymbol{\nabla}\cdot {\bf B}^{\tt a} \equiv 0\, ,
\end{equation}
which are the macroscopic Maxwell equations. We must look to the action to find the constitutive relations.

Ignoring surface terms, we may rewrite the action (\ref{AtAaction}) as \cite{Deser:1997gq,Schwarz:1993vs}
\begin{equation}\label{AtA1}
I[{\bf A}^1, {\bf A}^2] = \int\! dt\int\! d^3\sigma \, \left\{-\frac12 \epsilon_{\tt ab}\,  {\bf E}^{\tt a}\cdot {\bf B}^{\tt b} - {\cal H}({\bf B}^1, {\bf B}^2)  \right\}\, .
\end{equation}
In addition to being invariant under the gauge transformations (\ref{gaugetrans}), this action is also invariant under the following gauge transformation with another $SO(2)$-doublet of scalar parameters $\phi^{\tt a}$:
\begin{equation}\label{Azerotrans}
A_0^{\tt a}\to A_0^{\tt a} + \phi^{\tt a}\, .
\end{equation}
This is because the scalar potentials contribute only to a surface term in the action, which means that the field equations are found from variation of the vector potentials. Because of the identities (\ref{fromdefs}), these field equations are equivalent to
\begin{equation}\label{Afeqs}
    \boldsymbol{\nabla}\times \left({\bf E}^{\tt a} + \epsilon^{\tt ab}
    \frac{\partial{\cal H}}{\partial {\bf B}^{\tt b}}\right) ={\bf 0}\, .
\end{equation}
The gauge invariance (\ref{Azerotrans}) means that the electric fields ${\bf E}^a$ are, in this context, only defined up to the addition of
the gradients $\boldsymbol{\nabla}\phi^a$, so the general solution of
the field equations (\ref{Afeqs}) is gauge-equivalent to
\begin{equation}\label{E=dH}
{\bf E}^a = -\epsilon^{ab}
\frac{\partial{\cal H}}{\partial {\bf B}^b}\, .
\end{equation}
These are the constitutive relations.

Finally, we observe that in this new notation, the conditions (\ref{Linvcon}) and (\ref{Dinvcon}) for
Lorentz and duality invariance, respectively, are now
\begin{equation}
    \epsilon_{\tt ab}\left( {\bf B}^{\tt a} \times {\bf B}^{\tt b} - \frac{\partial{\cal H}}{\partial {\bf B}^{\tt a}}\times \frac{\partial{\cal H}}{\partial {\bf B}^{\tt b}}\right) = {\bf 0} \, , \qquad
    \epsilon_{\tt ab}\left( {\bf B}^{\tt a}\times \frac{\partial{\cal H}}{\partial {\bf B}^{\tt b}}\right)  = 0\, .
\end{equation}

\subsection{6D chiral 2-form: Hamiltonian formulation}\label{sec:Ham6}

We now replace the $3$-vector potential ${\bf A}$ by an antisymmetric tensor potential $\bA$, with components $\{A_{ij}; i,j, = 1,\dots, 5\}$. The analogs of the
electric field ${\bf E}$ and magnetic-induction field ${\bf B}$ are antisymmetric tensors $\bE$, $\bB$, with components
\begin{equation}
E_{ij} = 2\partial_{[i }A_{j]0} + \dot A_{ij} \, , \qquad B^{ij} = \frac12\varepsilon^{ijklm} \partial_k A_{lm}  := (\boldsymbol{\nabla} \times{\bA})^{ij}  \, .
\end{equation}
For a chiral theory,  $\bB$ is also the variable canonically conjugate to $\bA$, and the constraint imposed by $A_{j0}$ is an identity, so the action analogous to (\ref{4Dact}) is
\begin{equation}\label{6Dact}
I= \int\! dt\int\! d^5 \sigma \left\{\frac14 \dot A_{ij} B^{ij}  - {\cal H}(\bB)  \right\}\, .
\end{equation}
The normalization of the first term  differs from that used in \cite{Townsend:2019ils} but is more convenient for current purposes.
It implies the Poisson-bracket relation
\begin{equation}
\left\{ B^{ij}(\boldsymbol{\sigma}), B^{kl}(\boldsymbol{\tilde\sigma}) \right\}_{PB} = \varepsilon^{ijklp} \partial_p \, \delta(\boldsymbol{\sigma} - \boldsymbol{\tilde\sigma}) \, .
\end{equation}
The field equation obtained by variation of $\bA$ is
\begin{equation}\label{Bfield-eq}
\dot\bB  =\boldsymbol{\nabla}\times \bH\,  ,
\qquad \bH \equiv  \frac{\partial {\cal H}}{\partial \bB}\, .
\end{equation}
The antisymmetric-tensor field $\bH$ is the 6D analog of the magnetic field ${\bf H}$ of non-linear 4D electrodynamics.

The field equation (\ref{Bfield-eq}) implies that
\begin{eqnarray}\label{dots}
\partial_t{\cal H} &=& \bfn \cdot  (\bH\times \bH)  \, , \nonumber \\
\partial_t (\bB\times \bB)_i &=& \partial_k \left[2 H_{ij}B^{jk}   - \delta_i^k \left( \tr(\bB\bH)  + 2{\cal H} \right)\right] \, ,
\end{eqnarray}
where\footnote{\label{f1} The different normalization from \cite{Townsend:2019ils} compensates for the
different normalization of the symplectic form defined by the phase-space action.}
\begin{equation}\label{BxB}
 (\bB\times \bB)_i = \frac18\varepsilon_{ijklm}B^{jk} B^{lm}\, .
\end{equation}
These equations imply that the following integrals are constants of the motion (if surface terms are assumed to vanish):
\begin{equation}\label{N-trans}
H= \int\! d^5\sigma\, {\cal H}\, , \qquad
\boldsymbol{P} = - \int\!d^5\sigma \,  (\bB\times\bB)\, .
\end{equation}
These are the Noether charges associated, respectively,  to the time translation and  space translation invariances of the action.
As a check on this interpretation, and the normalizations, one may verify that for any smooth  function  $f$ of $\bB$ satisfying the field equation
(\ref{Bfield-eq}), and constants $(\alpha^0,\alpha^i)$,
\begin{equation}\label{translations}
\left\{ f, \int\! d^5\sigma \left[ \alpha^0 {\cal H} + \alpha^i (\bB\times\bB)_i \right] \right\}_{PB} = \left[\alpha^0 \partial_t + \alpha^i\partial_i \right]f\, .
\end{equation}
 This shows that the integrals of ${\cal H}$ and $\bB\times\bB$ are the components of a  6-covector. If we normalize this co-vector such that the dual 6-vector
has the Noether charge $H$ as its time component then its 5-space component is the Noether charge $\boldsymbol{P}$, irrespective of the signature convention
chosen for the Minkowski metric relating 6-vectors to 6-covectors.  It then follows that  the Noether charge associated to rotational invariance is
\begin{equation}
J^{ij} = -2\int\! d^5\sigma \, \sigma^{[i} (\bB\times\bB)^{j]} \, ,
\end{equation}
which is time-independent, for appropriate boundary conditions, as a consequence of the second of equations (\ref{dots}).

For the subset of the chiral 2-form theories defined by ${\cal H}(s,p)$, those that are Lorentz invariant will have, as an additional Noether charge, the Lorentz boost generator
\begin{equation}
\boldsymbol{L} = t \boldsymbol{P}   - \int\! d^5\sigma\, \boldsymbol{\sigma} \, {\cal H}\, .
\end{equation}
A calculation, using the first of eqs. (\ref{dots}), yields
\begin{equation}
\dot{\boldsymbol{L}}  = \boldsymbol{P} + \int\! d^5\sigma\, (\bH\times \bH) \, .
\end{equation}
From the expression for $\boldsymbol{P}$ in (\ref{N-trans}) we see that $\dot{\boldsymbol{L}}={\bf 0}$ requires
\begin{equation}\label{HH=BB}
\bH \times\bH = \bB\times\bB\, .
\end{equation}

As a check on the interpretation of the 5-vector Noether charge $\boldsymbol{L}$, we may compute the Poisson brackets
of its components. Using (\ref{translations}), and ignoring surface terms, we find that
\begin{eqnarray}\label{Lalg}
\left\{ L^i,L^j\right\}_{PB} &=& \iint\! d^5\sigma d^5\tilde\sigma\  \sigma^i \tilde\sigma^j
\left\{{\cal H}(\boldsymbol{\sigma}), {\cal H}(\boldsymbol{\tilde\sigma})\right\}_{PB} \nonumber \\
&=&  \iint\! d^5\sigma d^5\tilde\sigma\  \sigma^i \tilde\sigma^j  \left[ M^k(\boldsymbol{\sigma}) +
M^k(\boldsymbol{\tilde\sigma})\right] \partial_k\,  \delta^5(\boldsymbol{\sigma} - \boldsymbol{\tilde\sigma})\nonumber \\
&=& \int d^5 \sigma \, (\sigma^i M^j - \sigma^j M^i) \, ,
\end{eqnarray}
where $\boldsymbol{M}= (\bH\times\bH)$.  Since $\bH\times\bH = \bB\times\bB$ whenever ${\bf L}$ is a conserved charge, we have
\begin{equation}
\left\{ L^i,L^j\right\}_{PB} = - J^{ij} \, ,
\end{equation}
as expected for a Lorentz boost.

We now aim to determine the implications of the Lorentz invariance condition (\ref{HH=BB}) for the Hamiltonian density.
To this end it is convenient to use the $SO(5)$ rotational invariance to bring $\bB$ to a skew-diagonal form (at any chosen
spacetime point); we then have
\begin{equation}\label{sb}
B_{12} = -B_{21} =b_1\, , \qquad B_{34} =-B_{43} = b_2\, .
\end{equation}
This tells us that there are only two independent rotational scalars that we can construct from $\bB$, and we may take them to be
\begin{eqnarray}\label{sandp1}
s &=& \frac12 |\bB|^2 \equiv \frac14B^{ij}B^{kl}\delta_{ik}\delta_{jl}\, , \nonumber \\
p &=& |\bB\times \bB| \equiv \sqrt{(\bB\times \bB)\cdot (\bB\times \bB)} \, .
\end{eqnarray}
In terms of the skew-eigenvalues of $\bB$, we have
\begin{equation}\label{sandp2}
s = \frac12(b_1^2 + b_2^2)\, , \qquad p = |b_1b_2| \, .
\end{equation}
If we choose ${\cal H}$ to be a function of these two rotation scalars then $\bH$ has components
\begin{equation}\label{Hij}
H_{ij} = ({\cal H}_s + 2s p^{-1} {\cal H}_p) B_{ij} + p^{-1}{\cal H}_p (B^3)_{ij}\, .
\end{equation}
As expected, we have $\bH=\bB$ for the free-field Hamiltonian density ${\cal H}=s$. We also have
\begin{equation}
\bH\times \bH = \left[{\cal H}_s^2 + 2s p^{-1}{\cal H}_s{\cal H}_p + {\cal H}_p^2 \right] (\bB\times \bB)\, .
\end{equation}
The condition (\ref{HH=BB}) for Lorentz invariance is therefore equivalent to
\begin{equation}\label{6Dlorentz}
{\cal H}_s^2 + \frac{2s}{p} {\cal H}_s{\cal H}_p + {\cal H}_p^2  =1\, .
\end{equation}
This is formally identical to the equation (\ref{PDEsp}) required for Lorentz invariance of the generic 4D electrodynamics,
and this is why we have used the same notation.

We have already observed in the 4D context that  the condition (\ref{6Dlorentz}) takes the simpler form
\begin{equation}\label{6DLorentz-eq}
{\cal H}_{\tt u}{\cal H}_{\tt v} =1\,
\end{equation}
in terms of the new variables
\begin{equation}\label{uv6D}
{\tt u}= \frac12( s+ \sqrt{s^2-p^2} )\, , \qquad {\tt v} = \frac12( s- \sqrt{s^2-p^2} )\, ,
\end{equation}
but these variables are now $SO(5)$-invariant rotation scalars; they remain well defined in this context because it follows from (\ref{sandp2}) that
\begin{equation}\label{sp-ineq}
s^2-p^2 = \frac14(b_1^2-b_2^2)^2 \ge0\, .
\end{equation}
The solution of (\ref{6DLorentz-eq}) that previously led to the BI Hamiltonian density of
(\ref{cHDBI=}) now yields the following Hamiltonian density for a chiral 2-form electrodynamics:
\begin{eqnarray}\label{M5H}
{\cal H} = \sqrt{T^2 + T|\bB|^2 + |\bB\times\bB|^2} -T\, .
\end{eqnarray}
This is the Hamiltonian density for the chiral 2-form on the 6D Minkowski worldvolume of a static planar M5-brane in an 11D
Minkowski vacuum \cite{Bergshoeff:1998vx,Townsend:2019ils}. Its $T\to \infty$ limit is the free theory of Henneaux-Teitelboim \cite{Henneaux:1988gg} but the $T\to 0$ limit yields
\begin{equation}\label{bbd6}
{\cal H}_{T=0} = |\bB\times\bB|\, .
\end{equation}
This defines the interacting conformal 6D chiral 2-form theory of \cite{Gibbons:2000ck,Townsend:2019ils} which is a 6D analog
of the 4D BB electrodynamics.

\subsubsection{Reduction/truncation to 4D}

The correspondence just established between 6D chiral 2-form electrodynamics and 4D duality invariant electrodynamics
can be understood very directly from the existence of a dimensional-reduction/truncation that takes any given
6D example into its corresponding 4D example. The dimensional reduction step proceeds by writing the 5-space coordinates
as $\{\sigma^\alpha, \sigma^4,\sigma^5; \alpha=1,2,3\}$
and taking all fields to depend only on the $\sigma^\alpha$. In this case, one finds that
\begin{eqnarray}\label{B6tA4}
B^{\alpha\beta} &=& \varepsilon^{\alpha\beta\gamma} \partial_\gamma A_{45} \, , \qquad B^{45} = \frac12 \varepsilon^{\alpha\beta\gamma} \partial_\alpha A_{\beta\gamma}\, , \nonumber \\
B^{\alpha 4} &=& \varepsilon^{\alpha\beta\gamma} \partial_\beta  \tilde A_\gamma :=  D^\alpha  \, , \qquad
B^{\alpha 5} = \varepsilon^{\alpha\beta\gamma} \partial_\beta A_\gamma :=  B^\alpha  \, ,
\end{eqnarray}
where
\begin{equation}
A_\alpha = -A_{\alpha 4}\, , \qquad
\tilde A_\alpha =  A_{\alpha 5} \, .
\end{equation}
The 6D rotation scalars $s$ and $p^2$  may now be written as
\begin{eqnarray}\label{spto4}
s \equiv \frac12|\bB|^2 &=&  \frac12 (|{\bf D}|^2 + |{\bf B}|^2) + \frac12\left[(B^{45})^2 + \frac12 B_{\alpha\beta} B^{\alpha\beta}\right]   \nonumber \\
p^2\equiv |\bB\times\bB|^2 &=& |{\bf D} \times {\bf B}|^2 + \frac18 (B^{45})^2\, B^{\alpha\beta}B_{\alpha\beta}\, .
\end{eqnarray}
As these expressions are at least quadratic in the variables $(A_{45}, A_{\alpha\beta})$, the truncation
\begin{equation}\label{truncate}
A_{45}=0\, , \qquad A_{\alpha\beta} =0\, ,
\end{equation}
is a consistent one in the sense that the full field equations with $A_{45}, A_{\alpha\beta}$ set to zero are equivalent to the equations obtained from the truncated action.

After this truncation we have
\begin{equation}
s\to   \frac12 (|{\bf D}|^2 + |{\bf B}|^2) \, , \qquad p \to  |{\bf D} \times {\bf B}|\, ,
\end{equation}
which are the expressions for the 4D rotation scalars $(s,p)$.
The 6D Hamiltonian density therefore becomes the 4D Hamiltonian density of a duality-invariant 1-form electrodynamics.
In addition,
\begin{equation}
    \frac14 \dot A_{ij} B^{ij} \to - \dot{\bf A} \cdot \boldsymbol{\nabla}\times \tilde {\bf A} + \, {\rm total\ derivative}\, ,
\end{equation}
which means that the 4D action is (\ref{AtAaction}), which we showed to be equivalent to the manifestly duality-invariant action (\ref{AtA1}). The off-shell duality invariance of the action obtained from the 6D Hamiltonian formulation is guaranteed by the fact that the electromagnetic duality group is the $SO(2)$ factor of the $SO(3)\times SO(2)$ subgroup of the $SO(5)$ rotation group preserved by the reduction/truncation.

\subsection{6D chiral 2-form: Perry-Schwarz formulation}\label{sec:PSform}

Although there is no standard Lagrangian formulation of chiral 2-form electrodynamics, there is an alternative canonical-type formulation in which the manifest symmetry is
5D Lorentz invariance. Let us take the 6D Minkowski metric to be
\begin{equation}
ds^2_6 = \eta_{mn}^{(5)}dx^m dx^n -(dy)^2 \qquad  (m=0,1,2,3,4),
\end{equation}
where $\eta^{(5)}= {\rm diag} (1,-1,-1,-1,-1)$. Let $A_{mn}$ be the 5D components of the 6D 2-form potential, and define
\begin{equation}
B^{mn} = \frac12 \varepsilon^{mnpqr} \partial_p A_{qr}\, .
\end{equation}
All fields still depend on $y$ in addition to the 5D Minkowski coordinates, and the generic Perry-Schwarz action for them takes the form \cite{Perry:1996mk}
\begin{equation}\label{PSact}
I_{PS}= \int\! dy \int\! d^5x \left\{ \frac14(\partial_y A_{mn})B^{mn}  -{\cal V}\right\}\, ,
\end{equation}
where ${\cal V}$ is a 5D Lorentz-scalar function of $B$. This is a kind of phase-space action in which the role of time is played by the space coordinate $y$, with  field equations that are first-order
in $\partial_y$ rather that $\partial_t$.

One possible basis for 5D Lorentz scalars is
\begin{equation}\label{QR}
Q = \frac14 B^{mn}B_{mn} \, , \qquad R = \sqrt{Q^2 - w^mw_m}\, ,
\end{equation}
where
\begin{equation}
w_m = \frac18\varepsilon_{mnpqr} B^{np} B^{qr} \, ,
\end{equation}
and the 5D Lorentz indices are raised or lowered with the 5D Minkowski metric; a useful identity (which is a consequence of the obvious identity $B^{[mn}B^{pq}B^{rs]} \equiv0$) is
\begin{equation}\label{wBid}
w_m B^{mn} \equiv 0\, \qquad (n=0,1,2,3,4).
\end{equation}
The variable $R$ is manifestly real when the one-form $w$ is spacelike. To see that this is still true
when $w$ is timelike we observe that in this case we may choose coordinates such that only $w_0$ is non-zero; the identity
(\ref{wBid}) then implies that $B^{0i}=0$ ($i=1,2,3,4$)  from which it follows that
\begin{equation}
Q= -\frac14 \tr\, \bB^2 \, , \qquad w_0^2= \frac14\left[ \tr\, \bB^4 - \frac12 (\tr\, \bB^2)^2\right]\, ,
\end{equation}
where (as in the previous subsection) $\bB$ is the $4\times4$ matrix with entries $B^{ij}$.
We then have
\begin{equation}\label{Q2positivity1}
Q^2 - w^mw_m = \frac{1}{16} (\tr\, \bB^2)^2 - w_0^2  =  \frac14 \tr \left[\bB^2 - \frac14 (\tr\, \bB^2)\bI_4 \right]^2 \ge0\, .
\end{equation}

The condition on ${\cal V}$ for 6D Lorentz invariance was found in \cite{Perry:1996mk} using
a different basis of 5D Lorentz scalars; the equivalent equation for ${\cal V}(Q,R)$ is
\begin{equation}\label{PSpde}
{\cal V}_Q^2  - {\cal V}_R^2 =1\,  .
\end{equation}
In terms of the new variables
\begin{equation}\label{UVvars}
U= \frac12(Q-R) \, , \qquad V= \frac12(Q+ R)\, ,
\end{equation}
the equation for 6D Lorentz invariance is \cite{Perry:1996mk}
\begin{equation}\label{PSeq}
{\cal V}_U {\cal V}_V =1\, .
\end{equation}
This equation  is formally equivalent to the equation (\ref{GibRash}) for the Lorentz invariant Lagrangian of a generic non-linear 4D electrodynamics to have EL equations that are
invariant under an $SO(2)$ electromagnetic duality. As we now explain,  this is not a coincidence.

\subsubsection{4D Reduction Redux}

If we dimensionally reduce the generic Perry-Schwarz action of (\ref{PSact}) from  6D  to 5D  by setting
\begin{equation}
\partial_y A_{mn} =0\, ,
\end{equation}
then we have a manifestly Lorentz invariant 5D theory with Lagrangian density
\begin{equation}
{\cal L}_{5D} = - {\cal V}\, .
\end{equation}
Let $\{x^\mu,x^4; \mu=0,1,2,3\}$ be the 5D Minkowski coordinates, and let us dimensionally reduce/truncate  to 4D by setting
\begin{equation}
\partial_4 A_{\mu 4}=0\, , \qquad A_{\mu\nu}=0\, .
\end{equation}
Then the only non-zero components of $B$  are
\begin{equation}\label{4DB1}
B^{\mu\nu} = \varepsilon^{\mu\nu\rho\sigma} \partial_\rho A_{\sigma 4}
= \frac12 \varepsilon^{\mu\nu\rho\sigma} F_{\rho\sigma} \, ,
\end{equation}
where $F=dA$ for $A= dx^\mu A_{\mu 4}$, and the only non-zero component of $w$ is
\begin{equation}
w_4 = \frac18 \varepsilon_{\mu\nu\rho\sigma} B^{\mu\nu} B^{\rho\sigma} = - \frac18 \varepsilon^{\mu\nu\rho\sigma} F_{\mu\nu} F_{\rho\sigma} \, .
\end{equation}
We now find that
\begin{eqnarray}\label{QRSP}
Q = S\, , \qquad R= \sqrt{S^2+P^2} \, ,
\end{eqnarray}
where $(S,P)$ are the standard 4D Lorentz scalars. It follows that $(U,V)$ of (\ref{UVvars}) are now the 4D variables introduced in (\ref{dual-inv}),  so the Perry-Schwarz equation
(\ref{PSeq}) for 6D Lorentz invariance of the generic 6D chiral 2-form electrodynamics (with 5D Lorentz invariance manifest) reduces to the Gibbons-Rasheed equation
(\ref{GibRash}) for electromagnetic duality invariance of the generic 4D 1-form electrodynamics, with Lagrangian density
\be\label{LSP}
{\cal L}_{4D}=-{\mathcal V(S,P)}.
\ee

\section{PST formulation}\label{sec:PST}
There exist several approaches to the construction of manifestly Lorentz-invariant duality-symmetric or chiral $p$-form actions; they all use additional fields of some kind  (e.g. \cite{Siegel:1983es,Kavalov:1986ki,McClain:1990sx,Wotzasek:1990zr,Bengtsson:1996fm,Pasti:1995tn,Pasti:1996vs,Berkovits:1996nq,Berkovits:1996rt,Belov:2006jd,Belov:2006xj,Sen:2015nph,Sen:2019qit,Mkrtchyan:2019opf,Townsend:2019koy,Andriolo:2020ykk,Vanichchapongjaroen:2020uns}). The PST formulation \cite{Pasti:1995tn,Pasti:1996vs} is particularly economical, since it uses a single auxiliary scalar field or, more precisely, a nowhere-null closed  one-form. The covariant PST construction allows for a straightforward coupling of chiral $p$-forms to gravity. In addition, it connects different non-manifestly Lorentz-invariant formulations \cite{Maznytsia:1998xw,Maznytsia:1998yr,Ko:2017tgo} (such as \cite{Zwanziger:1970hk} to \cite{Deser:1976iy} and \cite{Henneaux:1988gg} to \cite{Perry:1996mk}) and has led to some novel results; an example is an M5-brane action \cite{Bandos:1997ui,Aganagic:1997zq} of a type relevant to a more general construction that we discuss in this section and in sections  \ref{sec:HD} and \ref{sec:sect6}.

The PST formulation of chiral $2n$-form electrodynamics in $D=4n+2$ dimensions starts from a potential $2n$-form $A$ on spacetime with  $H=dA$ its $(2n+1)$-form field-strength, and a closed spacetime one form $v$, which may be timelike or spacelike. For the Minkowski spacetime background that we assume here, we have $v=da$ for scalar field $a$, and a PST gauge invariance that allows us to identify $a$ with the Minkowski time coordinate (if $v$ is timelike) or with a Minkowski space coordinate (if $v$ is spacelike). These two versions of the PST actions are off-shell inequivalent (i.e. not related by local field redefinitions or gauge transformations) but they are expected to be on-shell equivalent (i.e. have equivalent field equations) for the following reason\footnote{This may not apply in non-Minkowski backgrounds because of topological issues in the spacelike-$v$ PST formulation  \cite{Bandos:2014bva,Isono:2014bsa,Ko:2017tgo}, and in some cases that will be discussed later.}: the PST field equation reduces to a manifestly Lorentz invariant non-linear self-duality condition that involves $v$, but one expects to be able to rewrite this in a form that is both manifestly Lorentz invariant and $v$-independent because there is no obstacle to manifest Lorentz covariance at the level of field equations. This is known to be true for the linear chiral $p$-form theories and has been proved for the M5-brane equations,
with the help of their interpretation as Lorentz covariant superembeddings \cite{Howe:1996yn,Howe:1997fb,Howe:1997vn,Bandos:1997gm}.

We focus in this section on the 6D chiral 2-form theories. We choose standard Minkowski coordinates $\{x^M; M=0,1, \dots,5\}$ for the 6D Minkowski background, with a metric  of  `mostly minus'  signature. From the components $H_{MNP}$ of $H$, and $v_M$ of $v$, we construct the gauge-invariant anti-symmetric tensor density
\begin{equation}\label{BMN}
B^{MN} = -\frac16 \varepsilon^{MNPQRS} H_{PQR}\, v_S\, , \qquad (v_S=\partial_S a)\, .
\end{equation}
The general $6D$ chiral 2-form PST action has the following form:
\begin{equation}\label{L6d12}
S = \int d^6 x\left(\frac 1{4v^2} B^{MN} H_{MNP}\, v^P - {\cal V}\right)\, , \qquad (v^2=v^Sv_S),
\end{equation}
where ${\cal V}$ is a function of the two Lorentz scalars
\begin{equation}\label{Q1Q212}
 Q_1 = -\frac1{4v^2} B^{MN}B_{MN} \, , \qquad  Q_2 =  Q_1^2 +\frac 1{v^2}w^Mw_M \, ,
 \end{equation}
 with
\begin{equation}\label{omegaM}
w_M = -\frac1{8v^2} \varepsilon_{MNPQRS} B^{NP} B^{QR} v^S\, .
\end{equation}

Notice that $v_Mw^M \equiv 0$, so $w$ is non-timelike  if $v$ is timelike, but is unrestricted if $v$ is spacelike.
The variable  $Q_2$ is manifestly non-negative unless  either (i) $v$ is timelike and $w$ spacelike, or (ii) $v$ is
spacelike and $w$ timelike. However, in these cases we may construct the $6\times 6$ projector  matrix
\begin{equation}
{\tt P}^M{}_N = \delta^M{}_N - \frac{v^Mv_N}{v^2} - \frac{w^Mw_N}{w^2}\, .
\end{equation}
A calculation similar to the one leading to (\ref{Q2positivity1}) now yields
\begin{eqnarray}\label{Q2>0}
Q_1^2 + \frac{1}{v^2} w^Mw_M &=&  \frac{1}{4(v^2)^2} \left[ \tr B^4 - \frac14 (\tr B^2)^2 \right] \nonumber \\
&=& \frac{1}{4(v^2)^2}  \tr \left[ B^2 - \frac14 (\tr B^2) {\tt P} \right]^2 \ge0\, ,
\end{eqnarray}
where the second equality relies on the following properties of ${\tt P}$:
\begin{equation}
\tr\, {\tt P} =4\, , \qquad ({\tt P} B)^M{}_N = B^M{}_N\, \, .
\end{equation}
The second of these properties is a consequence of the identity $v_MB^{MN}\equiv 0$, because
(i) this implies that $\det B=0$, and (ii)
\begin{equation}
w_MB^{MN} \propto v^N\sqrt{\det B}\, ,
\end{equation}
which is  a consequence of the Schouten identity\footnote{That is, an identity equivalent to the obvious
over-antisymmetrization identity on seven indices.}
\begin{equation}
v^{[T} B^{PQ}B^{RS}B^{M]N} \equiv 6\,
v^N B^{[PQ}B^{RS}B^{MT]}\, .
\end{equation}
This concludes the proof that $Q_2\ge0$ in all cases.

Because the PST Lagrangian density (\ref{L6d12}) depends on the scalar field $a$ in addition to the chiral 2-form fields,
equivalence to the Hamiltonian or Perry-Schwarz formulations depends upon the possibility of a PST gauge invariance
that will allow $a$ to be set to some fixed function on spacetime. The  infinitesimal form of the PST gauge transformations
that allows this is
\begin{equation}\label{PSTinv}
\delta_\varphi  a=\varphi(x), \qquad \delta_\varphi A = \frac{ \varphi(x)}{v^2}\;  (i_vH_3 -{\cal V}_B)\, ,
\end{equation}
where ${\cal V}_B$ is the two form with components $\partial{\cal V}/\partial B^{MN}$.
However, this is a gauge invariance of the PST action with Lagrangian density (\ref{L6d12})
only if ${\cal V}$ satisfies  \cite{Buratti:2019guq}
\begin{equation}\label{Vpde1}
 \left(\frac{\partial{\cal V}}{\partial Q_1}\right)^2  -  4 Q_2 \left( \frac{\partial {\cal V}}{\partial Q_2}\right)^2=1 \, .
\end{equation}

We shall give more details about properties of the PST formulation (in particular the form of the PST field equations) in Sections \ref{sec:HD} and \ref{sec:sect6}.
Here we explain how the Hamiltonian and Perry-Schwarz formulations are recovered from the PST action by gauge fixing.

\subsection{Timelike {\it v}}\label{subsec:PST4D}

If $v$ is timelike then the PST gauge invariance allows us to choose $a=t\equiv x^0$.  The only non-zero components of $B$ and $w$ are now $\bB$
and $\bB\times\bB$, respectively, where $\bB$ is the 5-space antisymmetric tensor density of subsection \ref{sec:Ham6},  and the PST action becomes that
of \eqref{6Dact} with ${\cal V} = {\cal H}$. After making this PST gauge choice we also find that
\begin{equation}\label{Qstosp}
 Q_1 = -s \, , \qquad   Q_2=s^2-p^2 \, ,
\end{equation}
where $(s,p)$ are the 6D rotation scalars defined in (\ref{sandp1}). When written in terms of $(s,p)$ the condition (\ref{Vpde1}) on ${\cal V}$ is identical to (\ref{6Dlorentz}), which is the condition for the Hamiltonian density ${\cal H}(s,p)$ to define a (6D) Lorentz invariant theory.

To summarize, the PST action (\ref{L6d12}) for timelike $v$ is a `covariant' version of the Hamiltonian phase-space action; by using (\ref{Qstosp}) to rewrite (\ref{Vpde1}) as a PDE for ${\cal H}(s,p) = {\cal V}(Q_1,Q_2)$ one recovers the PDE (\ref{6Dlorentz}).

\subsubsection{4D reduction/truncation prior to gauge-fixing of $v$}

We have already seen that the phase-space action for the generic 4D Lorentz and duality invariant 1-form electrodynamics
may be found by a reduction/truncation of the phase-space action for the generic 6D chiral 2-form electrodynamics.
An analogous truncation/reduction may be carried out directly on the 6D PST action with timelike $v$, {\it prior to gauge fixing}. This yields the 4D PST action of \cite{Pasti:2012wv}, which reduces to the phase-space action (\ref{AtA1}) upon PST gauge fixing. Prior to this gauge fixing, both Lorentz and duality invariance are manifest. We present
a brief summary of this formulation of 4D nonlinear electrodynamics theories, as we shall be using its higher-dimensional extension \cite{Buratti:2019cbm} in Section \ref{sec:HD}.

The  duality doublets of 3-vector `electric' and `magnetic' fields of (\ref{AtA1}) are replaced by duality doublets of the 4-vector `electric' and `magnetic' fields
\begin{equation}\label{PST-EB}
E^{\tt a}_\mu = F^{\tt a}_{\mu\nu}\hat v^\nu \, , \qquad B^{\tt a}_\mu = \tilde F^{\tt a}_{\mu\nu} \hat v^\nu\, ,
\end{equation}
where $F^{\tt a}=dA^{\tt a}$ and $\tilde F^{\tt a}$ its Hodge dual, and
\begin{equation}\label{norm-v}
\hat v = v/\sqrt{v^2}\, , \qquad v= da\, .
\end{equation}
The action for these fields is\footnote{An example is the BI theory on the D3-brane considered in  \cite{Berman:1998va,Nurmagambetov:1998gp}.}
\begin{equation}\label{act2.0}
 I[A^1,A^2] = \int\! d^4x\,\left\{\frac12\varepsilon_{\tt ab}\, E^{\tt a}\cdot B^{\tt b} -{\cal H}(B^1,B^2)\right\} \, ,
\end{equation}
where
\be\label{ebc0}
E^{\tt a}\cdot B^{\tt b} =   E^{\tt a}_{\mu}B^{{\tt b}\, \mu} = E_0^{\tt a}B_0^{\tt b} - \sum_{i=1}^3
E_i^{\tt a} B_i^{\tt b} \, .
\ee
This action is invariant under the gauge transformation \cite{Pasti:1995tn}
\begin{equation}\label{PST2}
\delta A^{\tt a} = v \phi^{\tt a}\, ,
\end{equation}
where $\phi^{\tt a}$ is a duality doublet of  scalar parameters.
It is also invariant under the (PST) gauge transformation
\begin{equation}\label{PST1}
\delta a=\varphi(x), \qquad \delta A^{\tt a}=-\frac{ \varphi(x)}{\sqrt{v^2}}\;  \left(E^{\tt a} -\varepsilon^{\tt ab}\frac{ \partial{\cal H}}{\partial{B^{\tt b}}}\right)\, ,
\ee
for an arbitrary function $\varphi$. This allows the gauge choice $a=t$, which reduces the action to (\ref{AtA1}), as claimed\footnote{In verifying this, and statements below, it should be remembered that we use a Lorentz metric with `mostly-minus' signature, as in \eqref{ebc0}.}. It also reduces the gauge transformation (\ref{PST2}) to the gauge transformation (\ref{Azerotrans}) that leaves invariant the action (\ref{AtA1}).

If ${\cal H}$ is $SO(2)$-duality invariant then it must be some function of the
duality-invariant Lorentz scalars
\begin{equation}\label{PSTscalars}
 s= -\frac12 B^{\tt a}\cdot B^{\tt a}   \, , \qquad  p= \sqrt{q}\, ,
 \end{equation}
where $q$ is the $Sl(2;\bR)$-duality invariant
 \begin{equation}
q= \frac12 \epsilon_{\tt ac}\epsilon_{\tt bd} (B^{\tt a}\cdot B^{\tt b})(B^{\tt c}\cdot B^{\tt d})\,.
\end{equation}
The notation is motivated by the fact that in the PST gauge $v=dt$ we find, using the definitions of ${\bf B}^{\tt a}$ in (\ref{EBcorresp}),  that
\begin{equation}
    s \to \frac12(|{\bf D}|^2 + |{\bf B}|^2) \, , \qquad
    p \to  |{\bf D}\times {\bf B}|\, ,
\end{equation}
which are the independent rotation scalars introduced in subsection \ref{subsec:4D}. As we saw for the 6D case,  the
PDE (\ref{PDEsp}) that is required for Lorentz invariance of the 4D phase-space action with Hamiltonian density ${\cal H}(s,p)$ becomes
the condition for PST gauge-invariance of the PST action (\ref{act2.0}) when $(s,p)$ are re-interpreted as the duality-invariant Lorentz scalars (\ref{PSTscalars}). When rewritten as a PDE for ${\cal H}(s,q)$, this condition is
\be\label{PSEsq}
{\cal H}_s^2+4s {\cal H}_s{\cal H}_q+4q {\cal H}_q^2=1\, .
\ee

\subsection{Spacelike {\it v}} \label{spacev}

When $v$ is spacelike the PST gauge invariance allows us to set $a=y\equiv x^5$.  For this PST gauge choice the only non-zero components of $B$ and $w$ are
\begin{equation}
B^{mn} = \frac12 \varepsilon^{mnpqr}\partial_p A_{qr} \, , \qquad w_m =-\frac18 \varepsilon_{mnpqr} B^{np}B^{qr}\, ,
\end{equation}
and the Lagrangian density (\ref{L6d12}) becomes that of the Perry-Schwarz action (\ref{PSact}).
Also, the variables $( Q_1, Q_2)$ on which ${\cal V}$ depends are now
\begin{eqnarray}\label{Qspace1}
 Q_1  = \frac14 B^{mn} B_{mn} =Q \, , \qquad
 Q_2  =  Q^2 -  w^mw_m  =R^2\, ,
\end{eqnarray}
where $(Q,R)$ are the variables introduced in (\ref{QR}). Using this result to rewrite  (\ref{Vpde1}) as an equation
for ${\cal V}(Q,R)$ we recover (\ref{PSpde}).

To summarize, the PST action for spacelike $v$ is a `covariant' version of the Perry-Schwarz action.  Hence, its  $4D$ dimensional reduction/truncation with $v=dx^5$ yields the $4D$ Lagrangian \eqref{LSP}.

\subsection{Variant PST formulations}
In addition to the formulation which uses the one form $v=da$ considered above, there are other variants of the PST action. For instance, in $D=6$ one can dualize the 
the (timelike) one-form $v=da$  to a 5-form field-strength whose Hodge dual is a nowhere-zero spacelike one-form $u$;  this yields a `dual' PST formulation \cite{Maznytsia:1998yr,Ko:2017tgo}. Technically, this procedure amounts to performing the following replacements in the PST Lagrangian \eqref{L6d12}: $B_{MN} \to H_{MNL}u^L$ and $H_{MNL}v^L \to - B_{MN}$ with the timelike $v$ replaced by the spacelike $u$.
In this case the reduction/truncation to 4D yields, on setting $u=dx^5$, the Lagrangian density of \eqref{LSP}.
In contrast, for the gauge choice $u=dx$ for a non-compact coordinate $x$, reduction/truncation yields a non-linear generalization of the duality-symmetric electrodynamics of Zwanziger \cite{Zwanziger:1970hk}. An example is the dual manifestly duality-symmetric Born-Infeld-like action on the D3-brane \cite{Lee:2013ewa,Vanichchapongjaroen:2017zuh} which can be found by dimensional reduction/truncation of a corresponding variant of the M5-brane action \cite{Ko:2016cpw,Ko:2017tgo}. Yet another form of the M5-brane action,
with a triplet of auxiliary closed one-forms, was constructed in \cite{Ko:2013dka}. It is related to an effective gauge field theory for multiple M2-branes with volume preserving diffeomorphisms of an `internal' 3-manifold as its gauge group \cite{Ho:2008nn,Ho:2008ve,Bandos:2008fr,Bandos:2008jv,Pasti:2009xc}.

\section{New examples in 4D and 6D}\label{newD46}

So far, we have exhibited a correspondence between any given Lorentz-invariant 6D theory of chiral 2-form electrodynamics and an associated  Lorentz and duality invariant 4D theory of 1-form electrodynamics. Within either the Hamiltonian or Lagrangian formulation, the correspondence comes about because the Hamiltonian/Lagrangian density is a function of two variables subject to a non-linear first-order PDE, which is formally the same in both cases for a particular choice of bases for the two sets of two independent variables. Moreover, the basis of independent variables can be chosen such that the PDE is also formally the same for both the Hamiltonian and
Lagrangian formulations (if, for 6D, we view as ``Lagrangian'' the formulation of \cite{Perry:1996mk} with manifest 5D Lorentz invariance). Thus, remarkably, the task of constructing these 4D/6D theories reduces to the solution of a single PDE for a function of two variables, where only the interpretation of variables distinguishes between 4D/6D and
Hamiltonian/Lagrangian.

One form of this `universal' PDE is (\ref{6DLorentz-eq}) where the dependent variable is the Hamiltonian density and
the independent variables $({\tt u,v})$ are (4D or 6D) rotation scalars. We have already discussed the solution that yields
Born-Infeld electrodynamics in the 4D context; as it applies equally in 6D, and as both have brane interpretations within String/M-theory, we shall call it the `D3/M5' solution.

Here we focus on  the one-parameter extension of the D3/M5 solution that we found recently in the context of 4D electrodynamics \cite{Bandos:2020jsw} by considering a Hamiltonian density of the form
\begin{equation}\label{HK}
{\cal H}({\tt u},{\tt v}) = \sqrt{K({\tt u},{\tt v})}
+ {\rm constant} \, .
\end{equation}
The `universal' PDE  (\ref{6DLorentz-eq}) satisfied by ${\cal H}$ becomes the following PDE for $K$:
\begin{equation}\label{Keq}
K_{\tt u} K_{\tt v} = 4K \, .
\end{equation}
An obvious ansatz for $K$ is the generic quadratic function
\begin{equation}
K= c+ b_1 {\tt u}+ b_2 {\tt v} + a_{11}{\tt u}^2 + 2a_{12} {\tt u}{\tt v} + a_{22} {\tt v}^2\, .
\end{equation}
This is a solution of (\ref{Keq}) provided that
\begin{equation}
a_{11}(a_{12} -1)=0\, , \qquad a_{22}(a_{12} -1)=0\, , \qquad  a_{11} a_{22} + a_{12}^2 = 2a_{12}\, ,
\end{equation}
and
\begin{equation}
(a_{12}-2)b_1 + a_{11} b_2 =0\, , \qquad (a_{12}-2)b_2 + a_{22} b_1 =0\, , \qquad b_1b_2  = 4c\, .
\end{equation}
There are two types of solution of these algebraic relations, according to whether we choose (i) $a_{12}=1$ or (ii) $a_{12}=2$, in which case $a_{11}=a_{22}=0$. For case (i), $K$ is a perfect square and this leads to
a Hamiltonian density linear in $({\tt u},{\tt v})$ that can be recovered as the weak-field limit of the Hamiltonian density resulting from case (ii). The algebraic relations for case (ii) determine $K$  in terms of a constant $T$  with dimensions of energy density,  and a dimensionless parameter $\gamma$.
Requiring ${\cal H}$ to be real for all values of $({\tt u},{\tt v})$ fixes the sign of $T$, and requiring it to be zero
in the vacuum fixes the arbitrary additive constant. This results in the Hamiltonian density
\begin{equation}\label{seconds}
{\cal H}= \sqrt{ T^2 + 2T(e^{-\gamma}{\tt u} + e^{\gamma}{\tt v}) + 4{\tt uv}} -T \, ,
\end{equation}
where $\gamma \in \mathbb R$ is a numerical parameter.

For $\gamma=0$ we recover the D3/M5 result. The strong-field  ($T\to0$) limit ${\cal H}=2\sqrt{\tt uv}$ is the same as that of the D3/M5 solution, see \eqref{bbham}, but the weak-field ($T\to\infty$) limit
still involves the parameter $\gamma$:
\begin{eqnarray}\label{Tinf}
{\cal H}|_{T=\infty} &=& e^{-\gamma} {\tt u} + e^{\gamma} {\tt v} \, \nonumber \\
&=& (\cosh\gamma) s - (\sinh\gamma) \sqrt{s^2-p^2}\, .
\end{eqnarray}
This Hamiltonian density is manifestly non-negative; it is always real since (as we have already seen) $s^2-p^2\ge0$ for either the 4D or 6D interpretation of the variables $(s,p)$.
The 4D interpretation yields an interacting extension of Maxwell's equations that preserves both electromagnetic duality and conformal invariance \cite{Bandos:2020jsw}. We briefly review this result below, and then consider the analogous 6D chiral 2-form electrodynamics.

\subsection{ModMax electrodynamics}

Using the 4D interpretation for the parameters $({\tt u},{\tt v})$, and their relation to the rotation scalars $(s,p)$, the Hamiltonian density of (\ref{seconds}) is found to be
\begin{equation}\label{BIgammaH}
 {\cal H}_{(T)} = \sqrt{T^2 +2T\left[(\cosh\gamma)s - (\sinh\gamma)\sqrt{s^2-p^2}\right]+ p^2} -T\,.
\end{equation}
This is the one-parameter generalization of the BI Hamiltonian density found in \cite{Bandos:2020jsw}.
Its strong-field ($T\to0$) limit yields the Hamiltonian density \eqref{bbham} of Bialynicki-Birula electrodynamics. Its
weak-field ($T\to\infty$) limit is (\ref{Tinf}) with the 4D interpretation of $(s,p)$; i.e.
\begin{equation}\label{HamDB}
{\cal H} = \frac12 (\cosh\gamma)( |{\bf D}|^2 +|{\bf B}|^2) -  \frac12(\sinh\gamma)\sqrt{ (|{\bf D}|^2 +|{\bf B}|^2)^2 - 4|{\bf D}\times {\bf B}|^2} \, .
\end{equation}
The Maxwell Hamiltonian density is recovered for $\gamma=0$, while for $\gamma>0$ one gets the one-parameter modification of  Maxwell electrodynamics called  ``ModMax'' electrodynamics in \cite{Bandos:2020jsw}.
The  ModMax Hamiltonian field equations are
\begin{equation}
\dot {\bf B} = - \boldsymbol{\nabla}\times \left[{\cal A}_- \, {\bf D} - {\cal C} {\bf B}\right] \, , \qquad
\dot{\bf D} = \boldsymbol{\nabla}\times \left[{\cal A}_+ {\bf B} - {\cal C} {\bf D}\right]\, .
\end{equation}
for coefficient functions
\begin{equation}
{\cal A}_\pm = \cosh\gamma \pm \sinh\gamma \cos\Theta \, , \qquad {\cal C}= \sinh\gamma \sin\Theta\, ,
\end{equation}
where the angular variable $\Theta$ is most simply defined in terms of the rotation scalars $(\xi,\eta)$
of (\ref{BBvars}) by
\begin{equation}\label{defTh}
\left(\xi,\eta\right) = \sqrt{\eta^2+\xi^2}\, \left(\cos\Theta,\sin\Theta\right)\, .
\end{equation}

The Legendre transform of ${\cal H}_{(T)}$ of (\ref{BIgammaH}) yields the
Lagrangian density\footnote{This does  not appear in \cite{Hatsuda:1999ys} because only solutions of the PDE \eqref{PDESP}
that yield Maxwell electrodynamics in the weak-field limit were considered.}
\begin{equation}\label{BIgammaL}
  {\cal L}_{(T)} = T - \sqrt{T^2 -2T\left[(\cosh\gamma)S +  (\sinh\gamma)\sqrt{S^2 +P^2}\right] - P^2}\, .
\end{equation}
This result is most easily found using methods explained in Section {\ref{sec:HD}. It reduces to the usual BI Lagrangian density for $\gamma=0$, and in the $T\to\infty$ limit it reduces to the ModMax Lagrangian density
obtained in \cite{Bandos:2020jsw} by Legendre transform of the ModMax Hamiltonian density (\ref{HamDB}):
\be\label{LMM}
\mathcal L=-\frac12 (\cosh\gamma)( |{\bf E}|^2 -|{\bf B}|^2) + \frac12(\sinh\gamma)\sqrt{ (|{\bf E}|^2 -|{\bf B}|^2)^2 + 4|{\bf E}\cdot {\bf B}|^2}\,.
\ee
By construction, this Lagrangian defines a duality-invariant 4D electrodynamics, and one may verify that
it satisfies the duality-invariance condition \eqref{PDESP}; see also \cite{Kosyakov:2020wxv}.

We refer the reader to \cite{Bandos:2020jsw} for more details of properties of the ModMax theory, but we mention here that $\gamma\ge0$ is necessary to eliminate the possibility of superluminal propagation of small-amplitude fluctuations about a background solution of constant electric/magnetic fields, and that there are exact Maxwell-like plane-wave solutions for $\gamma\ge0$. See also \cite{Flores-Alfonso:2020euz,Bordo:2020aoz,Flores-Alfonso:2020nnd,Amirabi:2020mzv} for recent studies of ModMax effects on self-gravitating solutions in General Relativity.

An alternative to using a Legendre transform to relate 4D Hamiltonian and Lagrangian densities, is to obtain both from an analogous 6D chiral 2-form theory by reduction/truncation, so we now turn to this 6D theory.

\subsection{New chiral 2-form theories}

Using the 6D interpretation for the parameters $({\tt u},{\tt v})$, and their relation to the 6D rotation scalars $(s,p)$, the Hamiltonian
density (\ref{seconds}) is formally the same as it was for 4D, i.e. (\ref{BIgammaH}). Using (\ref{sandp1}); i.e. $s= \frac12 |\bB|^2$ and $p= |\bB \times\bB|$,  we arrive at the 6D Hamiltonian density
\be
{\cal H} = \sqrt{T^2 + T \left[(\cosh\gamma)|\bB|^2  - (\sinh\gamma) \sqrt{ |\bB|^4 -4|\bB\times\bB|^2} \right] + |\bB\times \bB|^2} -T\, .
\ee
This reduces to (\ref{M5H}) for $\gamma=0$, and the strong-field ($T\to0$) limit is again (\ref{bbd6}) irrespective of the value of $\gamma$.
However, the weak-field ($T\to\infty$) limit is
\begin{equation}\label{cH=conf6}
{\cal H}_{T=\infty}  = \frac12 (\cosh\gamma) |\bB|^2 - \frac12 (\sinh\gamma) \sqrt{ |\bB|^4 - 4|{\bB\times\bB|^2}} \, .
\end{equation}
This defines a new interacting chiral 2-form electrodynamics; we shall show later that it is conformal invariant.

The field equation  is (\ref{Bfield-eq}) (i.e. $\dot\bB  =\boldsymbol{\nabla}\times \bH$) and for the weak-field limit we have
\begin{equation}
\bH_{T=\infty} = \left[\cosh\gamma - (\sinh\gamma) \frac{s}{\sqrt{s^2-p^2}} \right]\bB -
(\sinh\gamma)\, \frac{p}{\sqrt{s^2-p^2}}\,  {\bf n} \times \bB\,  ,
\end{equation}
where ${\bf n}$ is the unit 5-vector in the direction defined by $-\bB\times \bB$, and
\begin{equation}
  \left[{\bf n} \times \bB\right]^{ij} := \frac12 \varepsilon^{ijklm} n_kB_{lm} \, .
\end{equation}
The sign choice made for the unit 5-vector ${\bf n}$ ensures that  it becomes the unit 3-vector of (\ref{E-BB}) after reduction/truncaton to 4D.
As we saw earlier, this procedure replaces the 6D interpretation of the variables $(s,p)$ by their 4D interpretation, which means that
the 6D Hamiltonian density (\ref{cH=conf6}) becomes the 4D ModMax Hamiltonian density of (\ref{HamDB}).

\subsubsection{PST formulation}

Let us return to the PST Lagrangian (\ref{L6d12}). The solution of \eqref{Vpde1} corresponding
to \eqref{seconds} is
\begin{equation}\label{Vpsi1}
{\cal V}_{(T)}= \sqrt{T^2 -2T\left[(\cosh\gamma)Q_1 + (\sinh\gamma)\sqrt{Q_2}\right] +Q_1^2-Q_2} -T\, .
\end{equation}
In the $v=dt$ gauge, for which $(Q_1,Q_2)$ are given in terms of the $SO(5)$ rotation variables by (\ref{Qstosp}), and ${\cal V}={\cal H}$,
we may use (\ref{uv6D})  to rewrite ${\cal V}_{(T)}$ in terms of $({\tt u},{\tt v})$; the result is precisely the Hamiltonian density of (\ref{seconds}).

Besides the manifest Lorentz invariance, an advantage of the PST formulation is that we can
also choose the $v=dx^5$ gauge to arrive at the Perry-Schwarz formulation of the same 6D field theory.
In this case  the reduction/truncation to 4D described in Section \ref{sec:PSform} yields  the  Lagrangian density (\ref{BIgammaL}) for the generalized BI theory. This illustrates the fact that switching from timelike $v$ to spacelike $v$ in the PST formulation  effects a Legendre transform of the 4D theory obtained by the reduction/truncation procedure described earlier.

\subsubsection{Conformal invariance of the weak/strong-field limits}

A feature of both the strong-field and weak-field limits of (\ref{Vpsi1}) is that there is no dependence
on dimensionful parameters. This suggests that these limits yield conformal chiral 2-form electrodynamics
theories, and this is already known to be true for the `M5' case \cite{Gibbons:2000ck,Townsend:2019ils}. Here we present a general proof of conformal invariance based on the observation of Zumino \cite{Zumino:1970tu} that a theory defined for an arbitrary background spacetime metric is conformal invariant for a Minkowski background if its action for a general background depends on the background metric only through its conformal class; i.e. if it is Weyl invariant\footnote{This observation is a special case of a more general one \cite{Faci:2013kea}: Weyl invariance for a general background implies invariance under the diffeomorphisms generated by the conformal Killing vectors of any specific choice of background metric.}.

A potential difficulty with this idea is that the generalization from Minkowski to generic spacetime metric could violate essential gauge invariances. This difficulty does not arise in our case because PST gauge invariance survives the coupling to gravity. In any case, Zumino's argument does not really depend on the generalization to an arbitrary background metric: it suffices to consider the metric of the Minkowski background in arbitrary coordinates. Once we have an action of coordinate independent form, which will be the case if it is the integral of a scalar density, then Weyl invariance implies that the action depends only on the conformal class of the background  metric, and is therefore invariant under those diffeomorphisms generated by its conformal Killing vectors, whose algebra is the algebra of conformal isometries of the background (6D Minkowski in our case).

To apply this method we rewrite the  PST Lagrangian density (\ref{BMN}) in arbitrary coordinates, with $g$ as the Minkowski metric. The definition of $B$ in (\ref{BMN}) is unchanged  if we interpret $\varepsilon$ as the metric-independent alternating
tensor density of unit weight defined (in any coordinate system) by $\varepsilon^{012345}=1$. If we also interpret ${\cal V}$ in (\ref{L6d12}) as a scalar function of the scalars $(Q_1,Q_2)$ then the integrand of \eqref{L6d12} becomes
\begin{equation}\label{Lg}
{\cal L}_{PST} = \frac{1}{4v^2}g^{PQ} (B^{MN} H_{MNP}\, v_Q) - \sqrt{|\det g|}\, {\cal V}\, ,
\end{equation}
where
\begin{equation}
v^2 = g^{MN}v_Mv_N\, .
\end{equation}
Similarly, the scalars $(Q_1,Q_2)$ in arbitrary Minkowski coordinates are
\begin{eqnarray}
  Q_1 &=& - \frac{1}{4v^2 |\det g|}g_{MP}g_{NQ} B^{MN} B^{PQ}   \nonumber \\
  Q_2 &=& Q_1^2 + \frac{1}{v^2 |\det g|} g^{MN} w_M w_N \, ,
\end{eqnarray}
where\footnote{Here we use the metric-independent alternating tensor density of weight $-1$ defined, in any coordinate system, by $\varepsilon_{012345}=1$.}
\begin{equation}
w_M = - \frac{1}{8 v^2}\,  g^{ST} (\varepsilon_{MNPQRS}B^{NP}B^{QR}\, v_T)\, .
\end{equation}
The factors of $1/|\det g|$ ensure that $(Q_1,Q_2)$ are scalars rather than scalar densities.

Next, we consider the effect of the Weyl rescaling
\begin{equation}
g\to \Omega^2 g \qquad (\Rightarrow \ \sqrt{|\det g|} \to \Omega^6 \sqrt{|\det g|}) \, .
\end{equation}
This has no effect on the first term of (\ref{Lg}), but it leads to the following
rescaling of $Q_1$ and $\sqrt{Q_2}$:
\begin{equation}
Q_1 \to \Omega^{-6} Q_1\, , \qquad \sqrt{Q_2} \to \Omega^{-6}\sqrt{Q_2} \, .
\end{equation}
Notice that the factors of $\Omega^{-6}$ here are precisely what is needed to cancel the factor
of $\Omega^6$ coming from the $\sqrt{|\det g|}$ factor multiplying ${\cal V}$, so the second
term of (\ref{Lg}) will also be unaffected by the Weyl rescaling if and only if ${\cal V}$ is a homogeneous
function of degree one in the variables $Q_1$ and $\sqrt{Q_2}$.

We conclude that the weak-field and strong-field limits yield conformal theories of
chiral 2-form electrodynamics. It is also true that these limits exhaust the possibilities
for conformal chiral 2-form electrodynamics. The reason is that conformal invariance requires ${\cal V}$ to be a homogeneous function of first degree in both $Q_1$ and $\sqrt{Q_2}$, in addition to satisfying (\ref{Vpde1}), but this is equivalent to requiring the Hamiltonian density ${\cal H}(s,p)$ to be a homogeneous function of first degree in both arguments, in addition to satisfying (\ref{6Dlorentz}). As a mathematical problem this is identical  to the one already solved in \cite{Bandos:2020jsw} for 4D duality-invariant electrodynamics; the result in that case was that BB-electrodynamics and ModMax electrodynamics (including Maxwell) are the only possibilities.

A corollary of this result is that the pairing of 4D with 6D theories, of the type under discussion, is such that if one is conformal invariant then so is the other, and that all
conformal electrodynamics theories (duality invariant in 4D and chiral in 6D) occur in
4D/6D pairs.

\section{Higher dimensions: duality in {\it D=4n}}\label{sec:HD}

The study of non-linear $(2n-1)$-form electrodynamics in a Minkowski spacetime of dimension $D=4n$ for $n>1$, in particular the implications of an $SO(2)$ or  $Sl(2;\bR)$ duality invariance, was initiated by Gibbons and Rasheed \cite{Gibbons:1995cv}, who proposed an $n>1$ analog of the Born-Infeld Lagrangian, whose Hamiltonian form and strong-field limit were later  found by Chruscinski \cite{Chruscinski:2000zm}.

As explained for 4D in Section \ref{sec:PST}, it is possible to make both Lorentz invariance and duality invariances manifest via a PST-type action that involves
an additional non-null closed 1-form $v$ and a PST gauge invariance. Assuming that $v$ is timelike, as we shall do now, the Hamiltonian formulation is recovered on imposing the PST gauge $v=dt$.  A higher-dimensional generalisation of this action was used in \cite{Buratti:2019cbm} to investigate possibilities for self-interactions for $n>1$ within the context of a perturbative expansion about the linear theory; in particular, new duality-invariant quartic self-interactions were found.

In this framework one starts from an $SO(2)$ doublet of $(2n-1)$-form gauge potentials $\{A^{\tt a}; {\tt a}=1,2\}$ with $2n$-form field strengths $F^{\tt a} =dA^{\tt a}$, which are used to define the following
``electric'' and ``magnetic" $(2n-1)$-form fields
\be\label{be1}
E^{\tt a} = i_{\hat v} F^{\tt a}\, , \qquad B^{\tt a} = i_{\hat v} \widetilde F^{\tt a} \, ,
\ee
where $\hat v$ is here the vector-field dual to the normalized 1-form
of (\ref{norm-v}); i.e. $\hat v^2=1$, and $\widetilde F^{\tt a}$ is the Hodge dual of $F^{\tt a}$. The action for these fields, generalising (\ref{act2.0}), is
\begin{equation}\label{act2n1}
 I[A^1,A^2] = \int\! d^{4n}x\,\left\{\frac12\epsilon_{\tt ab}\, E^{\tt a}\cdot B^{\tt b} -{\cal H}(B^1,B^2)\right\} \, ,
\end{equation}
where
\be\label{ebc1}
E^{\tt a}\cdot B^{\tt b} = \frac{1}{(2n-1)!}\,  (E_{\mu_1\cdots\mu_{2n-1}})^{\tt a} (B^{\mu_1\cdots\mu_{2n-1}})^{\tt b} \,,
\ee
and ${\cal H}$ is a function of a basis for the independent $SO(2)$-invariant Lorentz scalars formed with the $(2n-1)$-forms $\{B^1,B^2\}$.

We shall find it convenient to use a notation in which the $k$ Lorentz indices of a $k$-form are denoted collectively by $[k]$. For example, in this notation the scalar product (\ref{ebc1}) becomes
\begin{equation}
E^{\tt a}\cdot B^{\tt b} =\frac{1}{(2n-1)!}E^{\tt a}_{[2n-1]}B^{{\tt b}\, [2n-1]}\,.
\end{equation}
Also, the partial derivatives of ${\cal H}$ are defined by
\begin{equation}\label{partial}
dH = \frac{\partial {\cal H}}{\partial B^{\tt a}} \cdot dB^{\tt a}  =
 \frac{1}{(2n-1)!}\left(\frac{\partial {\cal H}}{\partial B^{\tt a}_{[2n-1]}}\right) d B^{\tt a}_{[2n-1]}\, .
\end{equation}
Using this notation the variation of the Lagrangian density induced by variation of the $(2n-1)$-form potentials $A^{\tt a}$ is, omitting a total derivative,
\be\label{deltaLag}
\delta{\cal L} = \frac{1}{[(2n-1)!]^2}\delta A^{\tt c}_{[2n-1]'} \,
\epsilon_{\tt ca}\,
\varepsilon^{\mu\nu [2n-1]'[2n-1]}\partial_\mu \left[\Xi^{\tt a}_{[2n-1]}\hat v_\nu \right]\, , \ee
where the repeated indices of $[2n-1]$ and $[2n-1]'$ are summed over, and
\be\label{calF:=}
\Xi^{\tt a}_{[2n-1]} = E^{\tt a}_{[2n-1]} - \epsilon^{\tt ab}\frac{\partial{\cal H}}{\partial B^{{\tt b}\, [2n-1]}}\, .
\ee
The equation of motion is therefore
\be\label{eofm}
d [\Xi^{\tt a} \wedge \hat v]=0\, .
\ee
We remark that because $i_vB^{\tt a}_{2n-1}=0$ and ${\cal H}$ is a Lorentz invariant function of $B^{\tt a}_{2n-1}$,
we have
\be\label{mathcals}
i_v \Xi^{\tt a} =0\, .
\ee
Recalling that $dv=0$, it is evident from (\ref{deltaLag}) that the action (\ref{act2n1}) is invariant under the infinitesimal
transformation
\begin{equation}\label{PST2=2}
\delta_\phi\! A^{\tt a} = v\wedge\phi^{\tt a}\, ,
\end{equation}
for $(2n-2)$-form parameters $\phi^{\tt a}$; this is a generalization of the gauge invariance of (\ref{AtA1}) with gauge transformation (\ref{PST2}).  It may be used to `gauge away' the exact form arising in the  first-integral of (\ref{eofm}), which thereby becomes equivalent to the $2n$-form equation $\Xi^{\tt a}\wedge \hat v =0$. Because of (\ref{mathcals}) there is a further equivalence to $\Xi^{\tt a} =0$.
To summarize, the field equation for the action (\ref{act2n1}) can be trivially once-integrated, and this first integral is gauge equivalent to the equation\footnote{This non-linear generalization of a self-duality
condition on $F^{\tt a}$ is also a generalization to $n\ge1$ of the $n=1$ constitutive relations in the form given in (\ref{E=dH}).}
\begin{equation}\label{gensd}
\Xi^{\tt a} = E^{\tt a} - \epsilon^{\tt ab} \frac{\partial {\cal H}}{\partial B^{\tt b}}=0\, .
\end{equation}
Generically, this is an equation not only for the field-strength $2n$-forms $F^{\tt a}$ but also for the PST scalar $a$ appearing through its derivative $v=da$,
but there are special cases for which the action is invariant under the
following 'first' PST gauge transformation \eqref{PST1} (analogous to \eqref{PSTinv}):
\be\label{PST11}
\delta_\varphi \,A^{\tt a}= - \,
\frac {\varphi} {\sqrt{(\partial a)^2}} \, \Xi^{\tt a} ,  \qquad \delta_\varphi a=\varphi\, .
\ee
The action has this PST gauge invariance provided ${\cal H}$ also satisfies \cite{Pasti:2012wv,Buratti:2019cbm}
\be\label{dcH2=B2}
\epsilon^{\tt ab}\varepsilon^{\mu\nu [.] [.]'}  \left( \frac {\partial {\cal H}}{\partial B^{{\tt a}[.]}} \, \frac {\partial {\cal H}}{\partial B^{{\tt b}[.]'}} - B^{\tt a}_{[.]}B^{\tt b}_{[.]'}
\right)=0\,\; .
\ee
When this condition is satisfied the PST scalar field $a$ is a gauge degree of freedom that has no effect on the dynamics. In these cases the field equation for $a$ is
\be\label{eoma}
\epsilon^{\tt ab}\varepsilon^{\mu\nu[2n-1]
[2n-1]'}\partial_{\mu}\left(\frac{\partial_{\nu}a}{(\partial a)^2}\Xi^{\tt a}_{[2n-1]}\Xi^{\tt b}_{[2n-1]'}\right)=0\, ,
\ee
which is identically satisfied when the gauge field equation \eqref{eofm} holds; this is the Noether identity guaranteed by Noether's second theorem.

\subsection{Generalized BI-type electrodynamics}\label{subsec:BI-type}

The condition \eqref{dcH2=B2} simplifies if one restricts ${\cal H}$ to be a function of the two Lorentz and $SO(2)$-duality invariants
\be\label{xy1}
s=-\frac{1}{2}\,B^{\tt a}\cdot B^{\tt a}, \qquad
  q = \frac12 \epsilon_{\tt ac}\epsilon_{\tt bd} (B^{\tt a}\cdot B^{\tt b})(B^{\tt c}\cdot B^{\tt d})\, ,
\ee
which generalize the $n=1$ duality invariant scalars of (\ref{PSTscalars}); in particular $q$ is invariant under the larger
$Sl(2;\bR)$ duality group.  In this case, \eqref{dcH2=B2} is equivalent to
\be\label{DI1}
{\cal H}_s^2+4s {\cal H}_s {\cal H}_q+4q {\cal H}_q^2=1\, ,  \qquad
\ee
which is identical to (\ref{PSEsq}); equivalently, to (\ref{PDEsp}) with $p=\sqrt{q}$.

Thus any solution of (\ref{DI1}) yields a $(2n-1)$-form electrodynamics not only for $n=1$ in $D=4$ but for any $n\ge1$ in $D=4n$. In particular, the solution
\be\label{Ham2n*}
{\cal H}=\sqrt{T^2+ 2T\left((\cosh\gamma)s-(\sinh\gamma)\sqrt{s^2-q}\right) +q}-T
\ee
for real parameter $\gamma$ yields a $D=4n$ generalization of the
generalized BI theory \eqref{BIgammaH}, which reduces for $\gamma=0$ to the Gibbons-Rasheed generalization of the BI theory \cite{Gibbons:1995cv}. The weak-field ($T\rightarrow\infty$) limit, for any $\gamma$, is a generalization of the ModMax electrodynamics, see the Hamiltonian \eqref{HamDB}, which reduces to
the free theory with ${\cal H}=s$ for $\gamma=0$.
The strong-field limit is, independent of the value of $\gamma$,
\begin{equation}\label{sqrtq}
{\cal H} = \sqrt{q}\, .
\end{equation}
In the PST gauge $v=dt$, this becomes the $Sl(2;\bR)$-invariant Hamiltonian density of Chruscinski's $D=4n$ generalization of BB electrodynamics \cite{Chruscinski:2000zm}.

In a Lagrangian formulation of a duality invariant theory we need only a {\it single} $(2n-1)$-form potential $A$ which will be identified with $A^2$. Thus $F :=dA \equiv dA^2$, and hence
\be
(E,B)=(E^2,B^2)\, .
\ee
Notice that this choice accords, in the gauge $v=dt$,  with the $n=1$ choice of (\ref{EBcorresp}).
The Lagrangian density ${\cal L}$ must be a Lorentz scalar function of $(E,B)$, which is restricted by $SO(2)$-duality invariance to satisfy a particular nonlinear differential equation \cite{Gibbons:1995cv}. This equation simplifies significantly if we restrict ${\cal L}$ to be a function of the Lorentz invariants
\be\label{alfabeta1}
\alpha =  -\frac{1}{2}\,(E\cdot E - B\cdot B)\, , \qquad  \beta=-\,\big(E\cdot B \big)^2\,.
\ee
For $n=1$ (i.e. $D=4$) we have
\be\label{spab1}
\alpha=S, \quad\beta =-P^2 \, ,
\ee
where $(S,P)$ are the Lorentz scalars of  \eqref{SPinv}, which provide a complete basis for any Lorentz scalar function; for $n>1$, functions of
$(\alpha,\beta)$ constitute a special class of Lorentz scalar functions. Given that ${\cal L}={\cal L}(\alpha,\beta)$, the condition for $SO(2)$-duality invariance of the
EL equations is
\be\label{GR1}
{\cal L}_\alpha^2+4\alpha {\cal L}_\alpha {\cal L}_\beta+4 \beta {\cal L}_\beta^2=1\, ,
\ee
which  is formally identical to \eqref{DI1}. A family of solutions is
\be\label{L2n*}
{\cal L}(\alpha,\beta)= T -
\sqrt{T^2 -2T\left[ (\cosh\gamma)\alpha+   (\sinh\gamma)\sqrt{\alpha^2-\beta}\right] +\beta} \, .
\ee
For the case of $n=1$ and $\gamma=0$, it is not difficult to show that in the PST gauge $v=dt$, the function ${\cal H}$ of
(\ref{Ham2n*}) is the BI Hamiltonian density, and the function ${\cal L}$ of (\ref{L2n*}) is the BI Lagrangian density, which is related to the Hamiltonian density by a Legendre transform with respect to ${\bf D}$; recall that $B^1$ is the 4-vector with components
$(0,{\bf D})$ for $n=1$ when $v=dt$.

More generally, we can define for all $n\ge1$,
\be\label{HamLag1}
{\cal L}(E,B)= \sup_{D} \left[-E\cdot D- {\cal H}(D,B)\right] \, ,
\ee
where $D=B^1$. We shall assume that ${\cal H}(D,B)$ is a strictly
convex function of $D$, so that $D$ is uniquely defined as a function of $E$ by
\be\label{e2b1}
E=-\frac{\partial {\cal H}(D,B)}{\partial D}\, .
\ee
In this case ${\cal L}$ will be a strictly convex function of $E$, which is a sufficient condition for the Legendre transform to be an involution. This guarantees equivalence of the Hamiltonian and Lagrangian field equations, and hence that the latter will be duality invariant if the former are duality invariant.  However, it is impractical to find an explicit solution of (\ref{e2b1}) for $D$ as a function of $(E,B)$ when $\gamma\ne0$.

The Legendre transform was found for the $n=1$ case in the weak-field limit (i.e. for the ModMax theory) by indirect means \cite{Bandos:2020jsw}; as expected, it is the weak-field limit of the
generalized BI Lagrangian density of (\ref{BIgammaL}).  We shall now show how this result can be extended beyond the weak-field limit to the full generalized BI theory for any $n\ge1$.

\subsection{Legendre transform redux}

We first consider the general case for which ${\cal L}= {\cal L}(\alpha,\beta)$ and ${\cal H}={\cal H}(s,q)$, assuming strict convexity; the results obtained
will allow us to relate the Lagrangian and Hamiltonian densities of the
generalized BI-type theories just discussed in the PST formulation. In the PST gauge $v=dt$, these functions become the conventional Lagrangian and Hamiltonian densities, so we are essentially implementing a Legendre transform but in a way that preserves manifest Lorentz invariance (although not manifest duality invariance).

Our starting point is the conjugate relation to (\ref{e2b1}):
\be\label{b12*}
 D=- \frac{\partial {\cal L}}{\partial E}={\cal L}_\alpha\, E+2\,B(E\cdot B){\cal L}_\beta.\qquad
\ee
Inserting this relation in  \eqref{xy1} and \eqref{HamLag1}  leads to  \cite{Buratti:2019cbm}
\bea
s&=&-(1+{\cal L}_\alpha^2-4\beta {\cal L}_\beta^2)\kappa
+\alpha {\cal L}_\alpha^2+2\beta {\cal L}_\alpha
{\cal L}_\beta\label{xxapp*}\\[4pt]
 q &=&
\left(\beta -4\alpha\kappa +4\kappa^2\right) {\cal L}_\alpha^2 \\ [4pt]
{\cal H}&=& 2(\alpha-\kappa){\cal L}_\alpha +2\beta {\cal L}_\beta
-{\cal L}\label{happ*}
 \eea
where
\be\label{kappa:=}
\kappa =\frac12(B\cdot B) \, .
\ee
Now we take the differentials of these equations; i.e.
\be \label{dhapp*}
d{\cal H}=2(\alpha-\kappa)dL_\alpha  +  2\beta dL_\beta  + L_\alpha d\alpha + L_\beta d\beta
-2 L_\alpha d\kappa\, ,
\ee
and analogous expressions for $(ds,dq)$. By substitution for $d{\cal H}$ and $(ds,dq)$ one may verify that
\begin{equation}
d{\cal H} = ({\cal L}_\alpha + 4\kappa {\cal L}_\beta)^{-1}
\left[ ds - \left(\frac{{\cal L}_\beta}{{\cal L}_\alpha}\right) dq \right]\, .
\end{equation}
The  duality invariance condition \eqref{GR1} satisfied by ${\cal L}$ has ensured the absence of a $d\kappa$ term on the right hand side.
From this result we deduce that
\begin{equation}\label{HsHq}
{\cal H}_s = ({\cal L}_\alpha + 4\kappa {\cal L}_\beta)^{-1} \, ,
\qquad {\cal H}_q = - \left(\frac{{\cal L}_\beta}{{\cal L}_\alpha}\right) ({\cal L}_\alpha + 4\kappa {\cal L}_\beta)^{-1}\, .
\end{equation}
One may also verify that
\be\label{pq*}
 \sqrt{\alpha^2-\beta}\,  {\cal L}_\beta = \sqrt{s^2-q}\,  {\cal H}_q \, ,
\ee
and that
\be\label{haml*}
{\cal H} + {\cal L} =\sqrt{(s^2-q){\cal H}_s^2+q} + \alpha {\cal L}_\alpha + 2\beta {\cal L}_\beta\, .
\ee
The sign of the square roots  in these equations may be checked by consideration of the free-field theory for which ${\cal H}=s$ and ${\cal L}=\alpha$. We note that
both $(s^2-q)$ and $(\alpha^2-\beta)$ are non-negative.

Equations (\ref{HsHq}) implicitly determine $(\alpha,\beta)$ as functions of $(s,q)$ and, given these functions, equations (\ref{DI1}) and (\ref{haml*}) uniquely determine ${\cal H}$.
Although we already know that the Legendre transform determines ${\cal H}$ implicitly, we now show how the new presentation of this fact allows us to obtain an explicit expression for the Hamiltonian density corresponding to the Lagrangian density of \eqref{L2n*}. For this case we find that
\be
\alpha {\cal L}_\alpha + 2\beta{\cal L}_\beta -{\cal L} = T\left[ \cosh\gamma \sqrt{1+ 4(\alpha^2-\beta){\cal L}_\beta^2} + 2\sinh\gamma \sqrt{\alpha^2-\beta}{\cal L}_\beta -1\right] \, ,
\ee
where the duality-invariance condition on ${\cal L}$, in the form
\be
({\cal L}_\alpha + 2\alpha{\cal L}_\beta)^2= 1+ 4(\alpha^2-\beta)L_\beta^2\, ,
\end{equation}
has been used to obtain the first term in the bracket on the right hand side. Now, using both \eqref{pq*} and \eqref{haml*} we deduce that
\be\label{H+T=}
\!\! {\cal H}+T =  \sqrt{(s^2-q){\cal H}_s^2+q} + T\left[ \cosh\gamma \sqrt{1+ 4(s^2-q){\cal H}_q^2} + 2\sinh\gamma \sqrt{s^2-q} {\cal H}_q\right],
\ee
which is a partial differential equation for ${\cal H}(s,q)$; taken together with \eqref{DI1}, we have a system of two simultaneous differential equations with the unique solution
\be\label{ham2n*}
{\cal H}=\sqrt{T^2+ 2T\left[-(\sinh\gamma)\sqrt{s^2-q}+(\cosh\gamma)s\right] +q}-T\, .
\ee
As expected, this is the Hamiltonian density of (\ref{Ham2n*}).


\section{Higher dimensions: chirality in {\it D=4n+2}}\label{sec:sect6}

A chiral $2n$-form electrodynamics is possible for $D=4n+2$, and a manifestly Lorentz invariant action can be found by a straightforward generalization of the 6D PST action of \eqref{L6d12}. We introduce a $(2n+1)$-form $H=d A$ and normalized PST 1-form $\hat v$, which we again assume to be timelike, and we again define `electric' and `magnetic' fields as
\begin{equation}
    {\cal E} = i_{\hat v}H \, , \qquad {\cal B}= i_{\hat v} \tilde H\, ,
\end{equation}
where $\tilde H$ is the Hodge dual of $H$.  The action is then
\begin{equation}\label{4n2}
S= \int\,d^{4n+2}x \left(\frac12 \,{\cal E}\cdot{\cal B} - {\cal V}({\cal B})\right) ,
\end{equation}
where we again use the notation
\begin{equation}
 {\cal E}\cdot {\cal B} =
 \frac{1}{(2n)!}{\cal E}_{M_1 \cdots M_{2n}} {\cal B}^{M_1 \cdots M_{2n}} \, ,
\end{equation}
and similarly for any other pair of (2n)-forms; Lorentz indices have been raised here with a Minkowski metric of `mostly-minus' signature. The PST gauge invariance of the action imposes the following condition on the `potential' function ${\cal V}({\cal B})$ \cite{Buratti:2019guq}:
\begin{equation}\label{PSTch}
\epsilon^{MN [.]\, [.]'} \left( \frac {\partial {\cal V}}{\partial {\cal B}^{[.]}} \, \frac {\partial {\cal V}}{\partial {\cal B}^{[.]'}} - {\cal B}_{[.]}{\cal B}_{[.]'} \right)=0\, ,
\end{equation}
where the notation is as in \eqref{partial} and \eqref{dcH2=B2}. In the $v=dt$ gauge ${\cal V}$ becomes the Hamiltonian density and this condition on it becomes the condition for Lorentz invariance of the phase-space action \cite{Townsend:2019koy}.

For $n=1$, the action \eqref{4n2} reduces to the 6D chiral 2-form action of \eqref{L6d12} when account is taken of the fact that the 2-form $B$ in that equation is $\sqrt{v^2} {\cal B}$. We have already discussed the possible choices for ${\cal V}$ in this $n=1$ case. For any $n>1$ there are two known possibilities:
\begin{itemize}

\item Free-field theory \cite{Henneaux:1988gg}:
\begin{equation}
    {\cal V} =  \frac12\,{\cal B}\cdot{\cal B} \, .
    \end{equation}
    This potential trivially satisfies \eqref{PSTch}.

\item Strong-field theory \cite{Townsend:2019koy}
\begin{equation}
    {\cal V} = \sqrt{-w^2}\, ,
\end{equation}
where $w$ is the $(4n+2)$-vector
\begin{equation}\label{wm}
    w_M = -\frac{1}{2[(2n)!]^2}\, \varepsilon_{MNM_1 \cdots M_{2n} N_1\cdots N_{2n}}{\cal B}^{M_1\cdots M_{2n}} {\cal B}^{N_1\cdots N_{2n}}\hat v^N\, .
    \end{equation}
 We have verified that this potential is also a solution of \eqref{PSTch}.

 \end{itemize}
For $n>1$ no other interacting theories are currently known, but some
restrictions on the possibilities have been found \cite{Buratti:2019guq}.

The formula (\ref{wm}) implies that
\begin{equation}
   - w^2= \frac{(4n)!}{4[(2n)!]^4} {\cal B}^{P_1\cdots P_{2n}} {\cal B}^{Q_1\cdots Q_{2n}}
    \left[ {\cal B}_{[P_1\cdots P_{2n}}{\cal B}_{Q_1\cdots Q_{2n}]}\right]\, .
\end{equation}
We give here relatively simple expressions for the $n=1,2$ cases:
\begin{eqnarray}
\!\!\!(n=1): &\! -w^2 =& \!\!\frac12 \,({\cal B}\cdot{\cal B})^2
    - \frac14 {\cal B}^{PQ} {\cal B}_{QR} {\cal B}^{RS} {\cal B}_{SP}\,, \\
\!\!\!(n=2): &\! -w^2 =& \!\! \frac12\, ({\cal B}\cdot{\cal B})^2 - \frac12({\cal B}^I {\cal B}^J)({\cal B}_I {\cal B}_J)
+ \frac{1}{16} ({\cal B}^{IJ}{\cal B} ^{KL})({\cal B}_{IJ} {\cal B}_{KL}) \, ,
\end{eqnarray}
where we have used the following notation for Lorentz tensors quadratic in ${\cal B}$:
\be
 ({\cal B}^I{\cal B}^J) = \frac{1}{3!} {\cal B}^{IPQR}{\cal B}^J{}_{PQR}\, , \qquad
({\cal B}^{IJ}{\cal B}^{KL}) = \frac12 {\cal B}^{IJPQ} {\cal B}^{KL}{}_{PQ}\, .
\ee

\subsection{Reduction to {\it D=4n}}

To perform the dimensional reduction/truncation to $D=4n$ we split the $D=4n+2$ Lorentz indices as follows
\be
M = (\mu,{\tt a})\, : \qquad \mu=0,1,\cdots, 4n\, ; \quad {\tt a} = 1,2\, .
\ee
We then set to zero all components of $H$ except $H_{\mu_1 \cdots \mu_{2n} {\tt a}}$, which we re-interpret as the components of a pair of $2n$-forms
$F_{\tt a}$, which we restrict to depend only on the  coordinates of the $D=4n$ Minkowski subspace.  We also restrict the closed PST 1-form $v$ to this  subspace, so
$v \to dx^\mu v_\mu$.  As a result
\be
{\cal E} \to E^{\tt a} \wedge dx_{\tt a} \, , \quad  {\cal B} \to \epsilon_{{\tt a}{\tt b}}\delta^{\tt bc} B^{\tt a} \wedge dx_{\tt c}\,, \qquad {\cal V}({\cal B}) \to {\cal H}(B^1,B^2) \, .
\ee
The Euclidean metric on the 2-dimensional compact space appears in the reduction/truncation of ${\cal B}$ because this was defined
using the Hodge dual. The fields $(E^{\tt a},B^{\tt b})$ on the $4n$-dimensional Minkowski spacetime are independent of this 2-metric,
so $\epsilon_{\tt ab} E^{\tt a}\cdot B^{\tt b}$ is an $Sl(2;\bR)$ invariant.  The action (\ref{4n2}) becomes the action (\ref{act2n1}),
and the constraint  (\ref{PSTch}) on ${\cal V}$ becomes the constraint (\ref{dcH2=B2})  on ${\cal H}$. In particular, ${\cal B} \cdot {\cal B} \to - B^{\tt a}\cdot B^{\tt a}$, which tells us that the truncation/reduction of the free chiral $2n$-for electrodynamics in $D=4n+2$ is the free duality invariant $(2n-1)$-form electrodynamics in $D=4n$.

\subsubsection{A new $D>4$ generalization of BB electrodynamics}

The above reduction/truncation takes $dx^Mw_M \to dx^\mu w_\mu$, where
\be
w_\mu = \frac{1}{2 [(2n-1)!]^2}\varepsilon_{{\tt a}{\tt b}} \, \hat v^\nu \varepsilon_{\mu\nu \rho_1\cdots \rho_{2n-1} \sigma_1\cdots \sigma_{2n-1}} (B^{\rho_1\cdots \rho_{2n-1}})^{\tt a}(B^{\sigma_1\cdots \sigma_{2n-1}})^{\tt b}
\ee
which yields
\be\label{w2:4n}
-w^2 = \frac{(4n-2)!}{4[(2n-1)!]^4}\epsilon^{\tt ab} \epsilon_{\tt cd} B_{\tt a}^{\mu_1 \cdots \mu_{2n-1}}
B_{\tt b}^{\nu_1\cdots\nu_{2n-1}}
\left( B^{\tt c}_{[\mu_1\cdots\mu_{2n-1}} B^{\tt d}_{\nu_1\cdots \nu_{2n-1}]} \right)
\ee
This Lorentz scalar is also manifestly $Sl(2;\bR)$ invariant. We give here simplified expressions
for $n=1,2$:
\begin{itemize}
\item $n=1$.
\be
-w^2 = B_{[{\tt a}}^\mu B_{{\tt b}]}^\nu \left( B^{\tt a}_\mu B^{\tt b}_\nu\right) = B^2D^2 - (B\cdot D)^2\, .
\ee
In the PST gauge $v=dt$ we have $\sqrt{-w^2}= |{\bf w}| = |{\bf D} \times {\bf B}|$, and hence BB electrodynamics.

\item $n=2$
\begin{eqnarray}\label{wsq}
-w^2 = q + q' \, ,
\end{eqnarray}
where $q$ is the $Sl(2;\bR)$ invariant of (\ref{xy1}),  and
\be
q' = -\frac18 \epsilon^{\tt ab}\epsilon_{\tt cd} (B_{\tt a}^{\mu\rho\sigma} B^{\tt c}_{\nu\rho\sigma})(B_{\tt b}^{\nu\eta\tau} B^{\tt d}_{\mu\eta\tau}) \, ,
\ee
which is another $Sl(2;\bR)$ invariant.
\end{itemize}
From the result for $n=2$ we see that the reduction/truncation of the `strong-field' chiral electrodynamics in $D=4n+2$
yields an $Sl(2;\bR)$-duality invariant electrodynamics in $D=8$ with
\be
{\cal H} = \sqrt{q+q'}\, .
\ee
This differs from the result (\ref{sqrtq}}) for the strong-field limit of the BI-type 8D duality-invariant electrodynamics
(and its one-parameter generalization) discussed in subsection \ref{subsec:BI-type}. Moreover, the expression (\ref{w2:4n})
shows that this difference will exist for all $n>1$ because the invariant $q$ is then just one term in the expansion of (\ref{w2:4n})
in $Sl(2;\bR)$ invariants.

Ths inequivalence for $n>1$ of the strong-field limit \cite{Chruscinski:2000zm} of the BI-type theories of \cite{Gibbons:1995cv} with
the `strong-field' theory obtained by  reduction/truncation of the chiral `strong-field' $D=4n+2$ theory \cite{Townsend:2019koy} is an indication that the strong-field limit of the BI-type theories can be `lifted' to $D=4n+2$ only for $n=1$. This would not be surprising because it is already known that the weak-field expansion of the BI-type theories is not the reduction/truncation of the weak-field expansion of any chiral $D=4n+2$ theory \cite{Buratti:2019guq}.

\section{Summary and discussion}\label{discussion}

We have explored multiple formulations of generic $SO(2)$-duality invariant
non-linear 4D electrodynamics theories and generic nonlinear 6D chiral 2-form
electrodynamics. In each case,  the generic model (whether 4D or 6D)
is determined by a function of two variables, and the condition for
both Lorentz invariance and duality (4D) or chirality (6D) requires this
function to satisfy a particular `universal' PDE.  Given a solution of this PDE,
the PST method allows the construction of an action for which both Lorentz invariance
and duality-invariance (4D) or chirality (6D) are manifest, but this depends on a new and
non-manifest (PST) gauge invariance; requiring this gauge invariance of the generic
PST-type action again leads to the `universal' PDE.

This 4D/6D universality is partly explained by the fact that any chiral 6D 2-form electrodynamics
theory contains, as a consistent reduction/truncation, a duality-invariant 4D electrodynamics theory.
Here we have shown that this process also maps the 6D `universal' PDE into  the 4D `universal' PDE.
This implies a one-to-one correspondence between the sets of 4D and 6D theories since both
are in one-to-one correspondence with the set of solutions to a single `4D/6D universal' PDE:
each solution yields both a duality invariant 4D electrodynamics and a 6D chiral 2-form electrodynamics
related by reduction/truncation. The general solution of this PDE is known 
but various physical constraints
(e.g. convexity, analyticity) mean that some solutions will have more physical relevance than others.

One well-studied solution yields both the 4D Born-Infeld theory, which arises in string-theory as the effective
dynamics on the worldvolume of a planar static D3-brane, and the chiral 2-form electrodynamics on the
worldvolume of a planar static M5-brane. We have called this the ``D3/M5 pair''; it is really a family of
paired 4D/6D theories parameterized by a constant with dimensions of energy density. This family includes
both a free-field limit, corresponding to weak fields with low energy density,  and an interacting conformal-invariant strong-field limit.
For 4D the strong-field limit is Bialynicki-Birula electrodynamics \cite{BialynickiBirula:1984tx,BialynickiBirula:1992qj}, which is the
unique 4D electrodynamics theory that is both Lorentz invariant and $Sl(2;\bR)$-duality invariant. The corresponding 6D conformal chiral 2-form electrodynamics was found by Gibbons and West \cite{Gibbons:2000ck} as a limit (and truncation) of the M5 Hamiltonian
density \cite{Bergshoeff:1998vx}, and its  6D Lorentz invariance was established in \cite{Townsend:2019ils}.

One issue that we have clarified here is the relation between the Lagrangian and Hamiltonian formulations of
Bialynicki-Birula electrodynamics. As shown in \cite{BialynickiBirula:1984tx}, the Lagrangian density found by Legendre transform of the Hamiltonian density is identically zero. Nevertheless, restrictions on the domain of this zero function contain the information that is required for reconstruction of the non-zero Hamiltonian density by a Legendre transform.
A corollary of this result is that an alternative `extended' Lagrangian, without restrictions on its domain, is a sum of
constraints imposed by Lagrange multipliers, as originally proposed in \cite{BialynickiBirula:1992qj}.  Many readers will be familiar with the possibility that a non-trivial dynamical system may have a zero canonical Hamiltonian;  Bialynicki-Birula electrodynamics
shows how a non-trivial dynamical system may have a zero canonical Lagrangian.

Another solution of the `universal' PDE yields, in 4D,  the new one-parameter generalization of the Born-Infeld
theory discussed in \cite{Bandos:2020jsw}; one of its  features  is that the weak-field and strong-field
limits exhaust the possibilities for conformal invariant 4D electrodynamics. The strong field limit is the same
as that of the  Born-Infeld theory; i.e. Bialynicki-Birula electrodynamics. Its weak-field limit is a new one-parameter
interacting generalization of Maxwell electrodynamics, which we called ``ModMax'' electrodynamics. This theory
does not have an analytic Lagrangian density but the Hamiltonian density is analytic within its `convexity domain'
\cite{Bandos:2020jsw}.  Here we have exhibited the corresponding family of 6D chiral 2-form electrodynamics; in this case the
weak-field limit is a conformal chiral 6D analog of the 4D ModMax  theory, which it contains as a consistent reduction/truncation. Together, these 4D/6D theories and their conformal limits constitute a `generalized' D3/M5 pair.

There is a natural extension of 4D duality invariant nonlinear electrodynamics to duality-invariant
$(2n-1)$-form nonlinear electrodynamics in a Minkowski spacetime of dimension $D=4n$, and a
generalization of Born-Infeld theory to these dimensions has been proposed by Gibbons and Rasheed \cite{Gibbons:1995cv}; its strong-field limit is
a generalization of Bialynicki-Birula electrodynamics that also has an enhanced $Sl(2;\bR)$-duality invariance.  It would be natural to suppose that this proposed generalization of
Born-Infeld electrodynamics is the reduction/truncation of a chiral  $2n$-form electrodynamics in $D=4n+2$ for $n>1$. For $n=2$, at least, we know that this supposition is false; this is because there is a unique quartic interaction in 10D and its reduction/truncation to 8D \cite{Buratti:2019guq} has a different form to that appearing in a weak-field expansion of the Gibbons-Rasheed theory \cite{Gibbons:1995cv}.

The only currently known 10D interacting chiral 4-form electrodynamics is the $n=2$ case of a class of conformal
invariant  chiral $2n$-form electrodynamics in $D=4n+2$ \cite{Townsend:2019koy}. The $n=1$ case is the strong-field limit of the `M5' chiral 2-form electrodynamics, so the $n=2$ case is a natural candidate for the strong-field limit of any proposed 10D generalization. Here we have shown that its truncation/reduction yields, as expected,  an 8D  conformal 3-form electrodynamics with an enhanced $Sl(2;\bR)$-duality invariance, but this is {\it not} the strong-field limit of the Gibbons-Rasheed theory.  A corollary of these results is that, for $n>1$, not all $(2n-1)$-form duality-invariant electrodynamics theories in $D=4n$ are obtainable by reduction/truncation from a chiral $2n$-form electrodynamics in $D=4n+2$. In other words, chirality in $D=4n+2$ implies
duality in $D=4n$ for all $n\ge1$ but the one-to-one correspondence between duality-invariant electrodynamics and chiral
electrodynamics in space-time with two more spatial dimensions is a special feature of the $n=1$ case; in fact, the only known higher-dimensional ($n>1$)
analog of the interacting  4D/6D pairs investigated in detail here is the one `strong-field' case found here by reduction/truncation of the one known conformal `strong-field' chiral electrodynamics in $D=4n+2$. It is possible that this really is a strong-field limit of
a new non-conformal $n>1$ pair but this remains to be determined.



\section*{\bf Note added} 

After this article was accepted for publication we became aware of  \cite{Denisov:2017qou,Denisova:2019lgr}, where properties of a generic conformal (but not necessarily duality-invariant) non-linear 4D electrodynamics are studied. Birefringence properties were found in   \cite{Denisov:2017qou}   with results for conformal theories in accord  with those of  \cite{Bandos:2020jsw}.   The coupling to gravity was considered in \cite{Denisova:2019lgr}, and it was shown that a certain condition on black hole charges restricts the form of the conformal electrodynamics Lagrangian density to a one-parameter 
extension of the Maxwell case. When expressed in terms of the Lorentz invariants $S$ and $P$ of our equation \eqref{SPinv},  
this restricted Lagrangian density is 
\begin{equation}
{\mathcal L}(c)= (1-c^2)\,S+c^2\,\sqrt{S^2+P^2} \qquad (c^2\leq 1)\, .  
\end{equation}
For $c^2\le \frac12$ we may write
\begin{equation}
c^2 = \frac12\left(1-e^{-2\gamma}\right) \qquad (\gamma\ge0) \, , 
\end{equation}
in which case 
\begin{equation}
{\cal L}(c) = e^{-\gamma}{\cal L}_{ModMax}(\gamma)\ . 
\end{equation}
The constant factor of $e^{-\gamma}$ can be removed by a rescaling of the fields, so the restricted conformal electrodynamics Lagrangian density of \cite{Denisova:2019lgr} is, for $2c^2\le 1$ and prior to coupling to gravity, equivalent to the duality-invariant ModMax electrodynamics of \cite{Bandos:2020jsw}, despite the fact that duality invariance was not assumed in 
\cite{Denisova:2019lgr}.

\section*{Acknowledgements}

IB and DS have been partially supported  by the Spanish MICINN/FEDER (ERDF EU)  grant PGC2018-095205-B-I00. The work of IB was also partially supported by the Basque Government Grant IT-979-16 and by the Basque Country University program UFI 11/55.
PKT has been partially supported by STFC consolidated grant ST/L000385/1.
\bigskip


\appendix


\providecommand{\href}[2]{#2}\begingroup\raggedright\endgroup

\end{document}